\PassOptionsToPackage{framemethod=tikz}{mdframed}
\documentclass[11pt,letterpaper,onecolumn]{arXiv-2605.19597v1/hunyuan_arxiv}

\graphicspath{{}{arXiv-2605.19597v1/}}

\usepackage[authoryear,sort&compress,round]{natbib}
\usepackage[most,breakable,skins]{tcolorbox}
\tcbuselibrary{skins,breakable}
\usepackage{microtype}
\usepackage{wrapfig}
\usepackage{listings}

\usepackage{booktabs}
\usepackage{tabularx}
\usepackage{array}
\usepackage{multirow}
\usepackage{tikz}
\usetikzlibrary{
    arrows.meta,
    positioning,
    fit,
    calc,
    backgrounds
}

\newtcblisting{toolcall}{
    listing only,
    listing options={
        basicstyle=\small\rmfamily,
        columns=fullflexible,
        keepspaces=true,
        breaklines=true,
        breakatwhitespace=false
    },
    colback=gray!5,
    colframe=gray!30,
    boxrule=0.4pt,
    arc=1pt,
    left=6pt,
    right=6pt,
    top=1pt,
    bottom=1pt,
    before skip=4pt,
    after skip=4pt
}

\let\cite\citep

\usepackage{amsmath,amsfonts,amssymb,amsthm}
\usepackage{booktabs,array,tabularx,multirow,ragged2e,xcolor}
\usepackage{tikz}
\usetikzlibrary{arrows.meta,positioning,shapes.geometric,fit,backgrounds,calc}
\usepackage[most]{tcolorbox}

\definecolor{exposeRed}{HTML}{B91C1C}
\definecolor{exploitAmber}{HTML}{B7791F}
\definecolor{misleadBlue}{HTML}{1D4ED8}
\definecolor{judgeBlue}{HTML}{1D4ED8}
\definecolor{softGray}{HTML}{F3F4F6}
\definecolor{borderGray}{HTML}{CBD5E1}
\definecolor{codeBg}{HTML}{F8FAFC}
\definecolor{codeFrame}{HTML}{D5DDE8}
\definecolor{codeTitleBg}{HTML}{EEF4FF}
\definecolor{codeTitleText}{HTML}{1E3A8A}
\definecolor{codeAccent}{HTML}{2563EB}
\definecolor{tableHeadBg}{HTML}{EAF1F8}
\definecolor{tableBandBg}{HTML}{F7FAFC}
\definecolor{tableGroupBg}{HTML}{F1F5F9}
\definecolor{tableRule}{HTML}{94A3B8}
\definecolor{tableText}{HTML}{1F2937}

\usepackage{enumitem}
\setlist[itemize]{leftmargin=*}
\setlist[enumerate]{leftmargin=*}

\theoremstyle{definition}

\newcommand{\tablefont}{\small\color{tableText}\renewcommand{\arraystretch}{1.12}}
\newcommand{\tableheader}{\rowcolor{tableHeadBg}}
\newcommand{\tablegroup}[2]{\rowcolor{tableGroupBg}\multicolumn{#1}{l}{\itshape #2}}

\newsavebox{\fullwidthbox}
\newlength{\fullwidthnarrow}
\newlength{\fullwidthwide}
\newcommand{\fullwidthtabular}[2]{%
  \begingroup
  \setlength{\tabcolsep}{0pt}%
  \sbox{\fullwidthbox}{\begin{tabular}{#1}#2\end{tabular}}%
  \setlength{\fullwidthnarrow}{\wd\fullwidthbox}%
  \setlength{\tabcolsep}{10pt}%
  \sbox{\fullwidthbox}{\begin{tabular}{#1}#2\end{tabular}}%
  \setlength{\fullwidthwide}{\wd\fullwidthbox}%
  \setlength{\tabcolsep}{\numexpr 655360*(\linewidth-\fullwidthnarrow)/(\fullwidthwide-\fullwidthnarrow)\relax sp}%
  \begin{tabular}{#1}#2\end{tabular}%
  \endgroup}
\AtBeginEnvironment{tabular}{\arrayrulecolor{tableRule}}
\AtBeginEnvironment{tabularx}{\arrayrulecolor{tableRule}}

\newtcblisting{boxedcode}[1][]{
  enhanced,
  listing only,
  listing engine=listings,
  breakable,
  width=0.88\textwidth,
  center,
  colback=codeBg,
  colframe=codeFrame,
  colbacktitle=codeTitleBg,
  coltitle=codeTitleText,
  fonttitle=\sffamily\bfseries\footnotesize,
  boxrule=0.45pt,
  arc=2pt,
  borderline west={2pt}{0pt}{codeAccent},
  left=9pt,
  right=9pt,
  top=7pt,
  bottom=7pt,
  before skip=8pt,
  after skip=10pt,
  listing options={
    basicstyle=\ttfamily\scriptsize\color{black!82},
    breaklines=true,
    columns=fullflexible,
    keepspaces=true,
    showstringspaces=false
  },
  #1
}

\AtBeginDocument{%
  \setlength{\parindent}{0pt}%
  \setlength{\parskip}{0.6\baselineskip plus 2pt}%
}

\makeatletter
\renewcommand{\abscontent}{}
\renewenvironment{abstract}
  {\begin{tcolorbox}[
      colback=gblue9!5,
      colframe=gblue9!5,
      boxrule=0pt,
      arc=6pt,
      left=8pt,right=8pt,
      top=6pt,bottom=6pt,
      breakable]
   \absfont}
  {\end{tcolorbox}\par\bigskip}
\makeatother

\usepackage{amsmath,amsfonts,bm}

\def\eqref#1{equation~\ref{#1}}

\def\1{\bm{1}}

\DeclareMathAlphabet{\mathsfit}{\encodingdefault}{\sfdefault}{m}{sl}
\SetMathAlphabet{\mathsfit}{bold}{\encodingdefault}{\sfdefault}{bx}{n}

\usepackage{fvextra}
\usepackage{hyperref}
\usepackage{longtable}
\usepackage{fontawesome5}
\usepackage{microtype}
\usepackage{booktabs}
\usepackage{pifont} 
\usepackage{multirow}
\usepackage{xspace}
\usepackage{color}
\usepackage{adjustbox}
\usepackage[edges]{forest}
\usepackage{amssymb}  
\usepackage{amsfonts}
\usepackage{tabularx}
\usepackage{tikz}
\usepackage{makecell}
\usepackage{caption}
\usepackage{subcaption}
\usepackage{amsmath}
\usepackage[utf8]{inputenc}
\usepackage[T1]{fontenc}
\usepackage{CJKutf8}

\DeclareUnicodeCharacter{2018}{`}%
\DeclareUnicodeCharacter{2019}{'}%
\DeclareUnicodeCharacter{201C}{``}%
\DeclareUnicodeCharacter{201D}{''}%
\DeclareUnicodeCharacter{2013}{--}%
\DeclareUnicodeCharacter{2014}{---}%
\DeclareUnicodeCharacter{2011}{-}%
\DeclareUnicodeCharacter{2026}{\ldots}%
\DeclareUnicodeCharacter{2022}{\textbullet}%
\DeclareUnicodeCharacter{25CF}{\textbullet}%
\DeclareUnicodeCharacter{2192}{\ensuremath{\rightarrow}}%
\DeclareUnicodeCharacter{00B7}{\textperiodcentered}%
\DeclareUnicodeCharacter{00BB}{\guillemotright}%
\DeclareUnicodeCharacter{00D7}{\ensuremath{\times}}%
\DeclareUnicodeCharacter{00B0}{\textdegree}%
\DeclareUnicodeCharacter{00B1}{\ensuremath{\pm}}%
\DeclareUnicodeCharacter{00F7}{\ensuremath{\div}}%
\DeclareUnicodeCharacter{00BA}{\textordmasculine}%
\DeclareUnicodeCharacter{00E9}{\'e}%
\DeclareUnicodeCharacter{200C}{}%

\usepackage{url}
\expandafter\def\expandafter\UrlBreaks\expandafter{\UrlBreaks\do\-}
\usepackage{nicefrac}
\usepackage{xcolor}
\usepackage[normalem]{ulem}
\usepackage[most]{tcolorbox}
\usepackage{colortbl}
\usepackage{tcolorbox}
\usepackage[tikz]{bclogo}
\usepackage{wrapfig}
\usepackage{float}
\usepackage{enumitem}
\usepackage{ifthen}
\ifdefined\directlua
  \IfFileExists{emoji.sty}{\usepackage{emoji}}{}%
\fi
\ifdefined\emoji\else
  \newcommand{\emojipenguin}{%
    \raisebox{-0.15\height}{\includegraphics[height=1.0em]{figures/penguin2.pdf}}%
  }%
  \providecommand{\emoji}[1]{\ifthenelse{\equal{#1}{penguin}}{\emojipenguin}{}}%
\fi

\usepackage{listings}
\usepackage{xparse}
\ExplSyntaxOn

\NewDocumentCommand{\gradientcell}{m}{
    \fp_set:Nn \l_tmpa_fp {#1}
    \fp_compare:nTF {\l_tmpa_fp > 20} {
        \cellcolor{green!60!yellow!\fp_eval:n{(\l_tmpa_fp-20)*2}!white}#1
    }{
        \fp_compare:nTF {\l_tmpa_fp > 10} {
            \cellcolor{yellow!60!red!\fp_eval:n{(\l_tmpa_fp-10)*3}!white}#1
        }{
            \cellcolor{red!\fp_eval:n{\l_tmpa_fp*5}!white}#1
        }
    }
}
\ExplSyntaxOff

\definecolor{deepred}{RGB}{230,101,101}

\definecolor{mygrey}{RGB}{105,105,105}

\usetikzlibrary{arrows.meta,positioning,fit,backgrounds,calc}

\renewcommand{\thefootnote}{\fnsymbol{footnote}}

\usepackage[scale=0.9]{tgheros}

\newcommand{\sysname}{Jarvis\xspace}
\newcommand{\sysaudio}{Jarvis-Audio\xspace}
\newcommand{\sysomni}{Jarvis-Omni\xspace}
\newcommand{\jev}{JEV\xspace}
\newcommand{\jevomni}{Jev-Omni\xspace}

\newcommand{\gate}{Native Gate\xspace}
\newcommand{\brain}{Decision Brain\xspace}
\newcommand{\ledger}{Task Ledger\xspace}
\newcommand{\delivery}{Delivery\xspace}
\newcommand{\memory}{Memory\xspace}
\newcommand{\timeline}{Evidence Timeline\xspace}
\newcommand{\belief}{Belief State\xspace}

\newcommand{\act}[1]{\textsf{#1}}
\newcommand{\op}[1]{\texttt{\small #1}}
\newcommand{\state}[1]{\textsf{\small #1}}

\definecolor{jBlue}{HTML}{174EA6}
\definecolor{jBlueMid}{HTML}{3B7DDD}
\definecolor{jBlueSoft}{HTML}{D6E4FA}
\definecolor{jBlueTint}{HTML}{EEF4FD}
\definecolor{jTeal}{HTML}{14998A}
\definecolor{jTealSoft}{HTML}{D3F0EC}
\definecolor{jAmber}{HTML}{E39B2D}
\definecolor{jAmberSoft}{HTML}{FBEBD2}
\definecolor{jCoral}{HTML}{D9534F}
\definecolor{jCoralSoft}{HTML}{F8DCDB}
\definecolor{jPlum}{HTML}{7C5CBF}
\definecolor{jPlumSoft}{HTML}{E9E2F6}
\definecolor{jSlate}{HTML}{64748B}
\definecolor{jSlateSoft}{HTML}{E2E8F0}
\definecolor{jInk}{HTML}{1F2937}
\definecolor{jPaper}{HTML}{F8FAFC}

\definecolor{oursRow}{HTML}{EAF2FE}
\definecolor{tbdText}{HTML}{94A3B8}

\definecolor{jBrand}{HTML}{0B57D0}
\definecolor{jMarkLight}{HTML}{B4D0FF}
\definecolor{jMarkDeep}{HTML}{5A3DBF}
\definecolor{jBarGray}{HTML}{CFD6DF}
\definecolor{jGander}{HTML}{2958EB}

\newcommand{\oursrow}{\rowcolor{oursRow}}

\newtcolorbox{takeaway}[1][]{
  enhanced, breakable,
  colback=jBlueTint, colframe=jBlueTint,
  borderline west={2.2pt}{0pt}{jBlue},
  arc=2pt, boxrule=0pt,
  left=9pt, right=9pt, top=5pt, bottom=5pt,
  before skip=6pt, after skip=8pt,
  fontupper=\small, #1
}

\newcommand{\optag}[2]{\tikz[baseline=(t.base)]\node[jtag=#1](t){#2};}
\newcommand{\numc}[1]{\tikz[baseline=(c.base)]\node[circle, draw=jInk!70, fill=white,
  inner sep=0.7pt, font=\sffamily\tiny\bfseries, text=jInk](c){#1};}

\tikzset{
  jnode/.style={rounded corners=3pt, draw=#1, fill=#1!12, line width=0.6pt,
                align=center, font=\small\sffamily, inner sep=4pt},
  jnode/.default=jBlue,
  jarrow/.style={-{Stealth[length=2.2mm]}, line width=0.7pt, draw=jInk!70},
  jdash/.style={-{Stealth[length=2mm]}, line width=0.6pt, draw=jSlate, dashed},
  jlabel/.style={font=\scriptsize\sffamily, text=jInk!75},
  jtag/.style={rounded corners=2pt, fill=#1, text=white, font=\scriptsize\sffamily\bfseries,
               inner xsep=3pt, inner ysep=1.5pt},
  jtag/.default=jBlue,
}

\pgfdeclareradialshading{jarvismark}{\pgfpointorigin}{%
  color(0bp)=(jMarkLight); color(10.5bp)=(jBlueMid); color(25bp)=(jMarkDeep); color(50bp)=(jMarkDeep)}
\tikzset{
  pics/jarvismark/.style={code={%
    \pgfmathsetmacro{\jmu}{#1/60}%
    \pgfmathsetmacro{\jmr}{2.2*#1/60}%
    \begin{scope}[x=\jmu pt, y=\jmu pt, shift={(-32,-32)}]
      \begin{scope}
        \clip (32,32) circle [radius=30];
        \shade[shading=jarvismark, overlay] (22.4,45.2) circle [radius=51];
      \end{scope}
      \path (2,2) rectangle (62,62);
      \foreach \bx/\bh in {15/9, 22.2/22, 29.8/34, 37.4/20, 44.6/9}
        \fill[white, rounded corners=\jmr pt] (\bx,32-\bh/2) rectangle ++(4.4,\bh);
    \end{scope}
  }},
  pics/jarvismark/.default=10pt,
}
\newcommand{\jarvismark}[1][10pt]{\tikz\pic{jarvismark=#1};}

\title{Jarvis: Coordination Deserves a Brain of Its Own\\in Agentic Interactive Models}

\makeatletter
\def\AB@authnote#1{}
\def\AB@affilnote#1{}
\renewcommand\AB@affilsepx{\protect\\ \protect\Affilfont}
\makeatother
\fancypagestyle{firststyle}{%
  \fancyhf{}%
  \fancyhead[L]{\includegraphics[height=20pt]{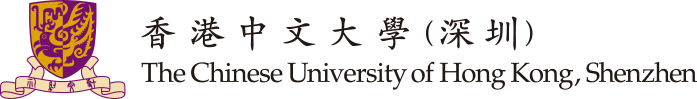}}%
  \fancyhead[R]{\includegraphics[height=13pt]{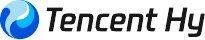}}%
}

\author{Yuxiang Wang\textsuperscript{1,2}, Shengbo Cai\textsuperscript{2}, Zhixian Zhao\textsuperscript{2}, Yingda Shen\textsuperscript{1}, Yuancheng Wang\textsuperscript{1},\\
\textbf{Xueyao Zhang\textsuperscript{1}, Liqiang Zhang\textsuperscript{2}, Teddy Sun\textsuperscript{2}, Steve Yves\textsuperscript{2}, Zhizheng Wu\textsuperscript{1,3}}\\
{\normalfont\small \textsuperscript{1}The Chinese University of Hong Kong, Shenzhen\quad \textsuperscript{2}Tencent Hunyuan\quad \textsuperscript{3}Amphion Technology Co., Ltd.}}

\begin{document}

\footnotetext[0]{\mbox{Email: \texttt{yuxiangwang1@link.cuhk.edu.cn}.}}

\maketitle

\vspace{-1.4em}

{\setlength{\topsep}{4pt}\begin{center}
    \raisebox{-0.15\height}{\jarvismark[15pt]}\hspace{7pt}%
    \textbf{Project Page: }\href{https://amphionteam.github.io/Jarvis-Demo/}{Jarvis Demos}%
    \hspace{12pt}%
    \raisebox{-0.12\height}{\textcolor{jInk}{\fontsize{15pt}{15pt}\selectfont\faGithub}}\hspace{7pt}%
    \href{https://github.com/JensenYX/Jarvis}{Code}%
\end{center}}

\begin{abstract}

Interactive models increasingly hand complex reasoning and tool use to a slow brain in the background, leaving a realtime front model free to sustain full-duplex dialogue. Yet the user may refine, cancel or replace a request while its task runs, so a plan or result that made sense a moment ago goes stale. Each new utterance must therefore be read against the work in flight, but the front model follows only the conversation and the back end sees only its own task. Coordination deserves a brain of its own, one that sees both. We introduce \textbf{\sysname}, a coordination harness where models judge and code enforces. Its typed decision model, JEV, reads each turn against the live tasks and picks among operations built from a versioned task record. Code checks every choice against the current state before applying it, so no decision can act on a task version the user has replaced. Jarvis also rechecks plans before they execute and results before they are spoken, and counts an answer as delivered only once the user has heard it in full. We build two instances of this design. Jarvis-Audio runs unchanged on three full-duplex speech models. Jarvis-Omni adds vision and rechecks the evidence behind an answer before and during playback, since a scene can change while nobody speaks. With the front model and back end held fixed, replacing Qwen Audio Agent with Jarvis raises $\tau$-Voice task success more than sixfold, from 29 to 192 of 278 tasks, and the gains hold on every benchmark and front model. Against published systems, Jarvis-Audio reaches 69\% Pass@1 on Full-Duplex-Bench v3 with an open 9B front model, versus 60\% for GPT-Realtime. These results suggest that how the two brains are coordinated matters as much as how strong they are.
\end{abstract}

\setcounter{footnote}{0}
\renewcommand{\thefootnote}{\arabic{footnote}}

\begingroup
\setlength{\intextsep}{3pt}\setlength{\abovecaptionskip}{6pt}
\begin{figure}[H]
\centering
\input{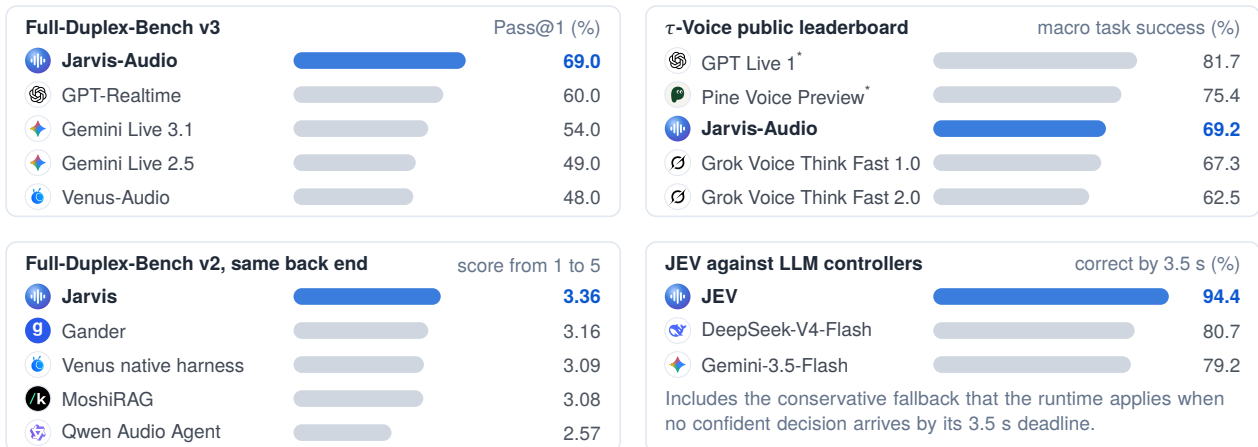}
\caption{Benchmark results for \sysname (blue) and other systems (gray). Top: the five highest published entries (Tables~\ref{tab:fdb3-published} and~\ref{tab:tau-published}). Bottom: runtimes on one \mbox{DeepSeek-V4-Flash} back end (Table~\ref{tab:paired}) and controllers with the same options (Table~\ref{tab:abl-jev}). $^*$Proprietary reasoning back end.}
\label{fig:headline}
\end{figure}
\endgroup

\clearpage
\suppressfloats[t]
\section{Introduction}
\label{sec:intro}

Full-duplex speech models have changed what it feels like to talk to a machine. Systems such as Moshi~\cite{defossez2024moshi}, MiniCPM-o~4.5~\cite{cui2026minicpm}, Realtime-Venus~\cite{venus2026}, GPT-Realtime~\cite{gptrealtime2025} and Gemini Live~\cite{gemini_live_2026} listen while they speak, take the floor at natural moments and stop when the user cuts in. As conversation becomes this natural, users expect these assistants to do real work as well, such as building a small party game, fixing a bug or remembering their usual snack budget. In games, they want a teammate that watches the match, suggests the next move and plays for them when asked. Such requests need tools, multi-step reasoning and long-term memory, more than a realtime model of a few billion parameters can supply while it keeps pace with the conversation.

Agentic interactive models close this gap with a division of labor. A realtime front model, the \emph{fast brain}, keeps the conversation going, while a \emph{slow brain}, either a larger model or a full agent harness, reasons and uses tools in the background~\cite{venus2026,qwen_audio_agent_2026,gander2026,Mind-paced-speaking,openai_gpt_live_system_card_2026}. This split mirrors the fast and slow modes of human thinking~\cite{kahneman2011thinking} and solves the capacity problem but creates a new one. Background work takes seconds or minutes, and the conversation does not wait. While a task runs, the user may refine, cancel or replace it, so a plan or result that made sense a moment ago goes stale.

\begin{figure}[t]
\centering
\begin{tikzpicture}[x=0.0535cm, y=1cm, font=\sffamily\scriptsize]
  \def\lanegap{0.52}
  \tikzset{
    vbar/.style={draw=jBlue, fill=jBlueSoft, rounded corners=1.2pt, line width=0.45pt},
    vlab/.style={font=\sffamily\tiny, text=jBlue!80!black},
    done/.style={circle, fill=jTeal, text=white, inner sep=0.6pt, font=\sffamily\tiny\bfseries},
    ask/.style={circle, fill=jAmber, text=white, inner sep=0.9pt, font=\sffamily\tiny\bfseries},
    gone/.style={circle, fill=jCoral, text=white, inner sep=0.8pt, font=\sffamily\tiny\bfseries},
    num/.style={circle, draw=jInk!70, fill=white, inner sep=0.7pt, font=\sffamily\tiny\bfseries, text=jInk},
    note/.style={font=\sffamily\tiny, text=jSlate},
  }

  \foreach \name/\y in {Game/4, Weather/3, News/2, Snacks/1, Python/0, Welcome sign/-1} {
    \node[anchor=east, text=jInk] at (22, \y*\lanegap) {\name};
    \draw[jSlateSoft, line width=0.4pt] (24, \y*\lanegap) -- (272, \y*\lanegap);
  }

  \foreach \t/\k in {28.9/1, 50.2/2, 69.2/3, 145.1/4, 161.8/5, 179.7/6, 232.6/7, 240.1/8} {
    \draw[jSlate!60, densely dashed, line width=0.35pt] (\t, 2.62) -- (\t, -0.82);
    \node[num] at (\t, 2.82) {\k};
  }
  \node[anchor=east, text=jInk, font=\sffamily\scriptsize\bfseries] at (22, 2.82) {User};

  \draw[vbar] (28.9, 4*\lanegap-0.15) rectangle (50.2, 4*\lanegap+0.15);
  \node[vlab] at (39.5, 4*\lanegap) {v1};
  \node[ask] at (42.8, 4*\lanegap+0.27) {?};
  \draw[vbar] (50.2, 4*\lanegap-0.15) rectangle (69.2, 4*\lanegap+0.15);
  \node[vlab] at (59.7, 4*\lanegap) {v2};
  \node[ask] at (60.2, 4*\lanegap+0.27) {?};
  \draw[vbar] (69.2, 4*\lanegap-0.15) rectangle (130.5, 4*\lanegap+0.15);
  \node[vlab] at (99.8, 4*\lanegap) {v3 \,builds the game};
  \node[done] at (130.5, 4*\lanegap) {\checkmark};
  \draw[vbar] (232.6, 4*\lanegap-0.15) rectangle (261.3, 4*\lanegap+0.15);
  \node[vlab] at (247, 4*\lanegap) {v4 \,light};
  \node[done] at (261.3, 4*\lanegap) {\checkmark};

  \draw[vbar] (28.9, 3*\lanegap-0.15) rectangle (84.7, 3*\lanegap+0.15);
  \node[vlab] at (56.8, 3*\lanegap) {v1 \,Shanghai, tonight};
  \node[done] at (84.7, 3*\lanegap) {\checkmark};
  \draw[vbar] (28.9, 2*\lanegap-0.15) rectangle (101.4, 2*\lanegap+0.15);
  \node[vlab] at (65.2, 2*\lanegap) {v1 \,two tech headlines};
  \node[done] at (101.4, 2*\lanegap) {\checkmark};

  \draw[vbar] (69.2, 1*\lanegap-0.15) rectangle (123.7, 1*\lanegap+0.15);
  \node[vlab] at (96.5, 1*\lanegap) {v1 \,usual \$40 from memory};
  \node[done] at (123.7, 1*\lanegap) {\checkmark};
  \draw[vbar] (179.7, 1*\lanegap-0.15) rectangle (215.0, 1*\lanegap+0.15);
  \node[vlab] at (197.4, 1*\lanegap) {v2 \,\$60 tonight};
  \node[done] at (215.0, 1*\lanegap) {\checkmark};

  \draw[vbar] (145.1, -0.15) rectangle (155.5, 0.15);
  \node[vlab] at (150.3, 0) {v1};
  \node[done] at (155.5, 0) {\checkmark};
  \draw[vbar] (161.8, -0.15) rectangle (172.5, 0.15);
  \node[vlab] at (167.1, 0) {v2};
  \node[done] at (172.5, 0) {\checkmark};

  \draw[draw=jCoral, fill=jCoralSoft, rounded corners=1.2pt, line width=0.45pt]
    (232.6, -1*\lanegap-0.15) rectangle (240.1, -1*\lanegap+0.15);
  \node[gone] at (240.1, -1*\lanegap) {$\times$};
  \node[note, anchor=west] at (244, -1*\lanegap) {only the sign is cancelled};

  \node[note, anchor=west, text=jPlum] at (218, 1*\lanegap) {usual \$40 kept};
  \node[note, anchor=east, text=jPlum] at (229.5, 4*\lanegap) {light theme saved as the new default};

  \draw[jInk!60, line width=0.4pt] (24, -0.86) -- (272, -0.86);
  \foreach \t in {30, 60, 90, 120, 150, 180, 210, 240, 270} {
    \draw[jInk!60, line width=0.4pt] (\t, -0.86) -- (\t, -0.92);
    \node[font=\sffamily\tiny, text=jInk!70, below] at (\t, -0.9) {\t\,s};
  }

  \node[done] at (30, -1.42) {\checkmark}; \node[note, anchor=west] at (33, -1.42) {delivered, playback confirmed};
  \node[ask] at (95, -1.42) {?}; \node[note, anchor=west] at (98, -1.42) {worker asks back};
  \node[gone] at (140, -1.42) {$\times$}; \node[note, anchor=west] at (143, -1.42) {cancelled};
  \draw[vbar] (178, -1.5) rectangle (190, -1.34); \node[note, anchor=west] at (192, -1.42) {one task version};
\end{tikzpicture}

\vspace{6pt}
{\sffamily\scriptsize\raggedright
\newcommand{\utt}[3]{\par\noindent\hangindent=1.25em\hangafter=1
  \makebox[1.25em][l]{\numc{#1}}\textit{#2}\par\noindent\hspace*{1.25em}#3\par\vspace{3pt}}
\begin{minipage}[t]{0.485\linewidth}
\utt{1}{``Build a star collecting game, 60 seconds a round. Check Shanghai's weather for tonight, and find two technology headlines.''}
  {\optag{jTeal}{NEW game} \optag{jTeal}{NEW weather} \optag{jTeal}{NEW news}}
\utt{2}{``Same phone, please. Two players playing together.''}
  {\optag{jBlue}{REFINE game v1}}
\utt{3}{``Use a shared score and let players respawn quickly. Also, make a snack plan using my usual budget.''}
  {\optag{jBlue}{REFINE game v2} \optag{jTeal}{NEW snacks}}
\utt{4}{``I pasted a little Python example. Why do both lists change?''}
  {\optag{jTeal}{NEW python}}
\end{minipage}\hfill
\begin{minipage}[t]{0.485\linewidth}
\utt{5}{``What is the smallest fix to make the lists independent?''}
  {\optag{jBlue}{REFINE python v1}}
\utt{6}{``For tonight, only sixty dollars is fine. Keep forty as my usual limit.''}
  {\optag{jBlue}{REFINE snacks v1} \optag{jPlum}{SAVE usual budget}}
\utt{7}{``Use a light theme from now on. Remember that as my new default, and update this game too. Also, make a separate full-screen welcome sign for the door.''}
  {\optag{jPlum}{SAVE light theme} \optag{jBlue}{REFINE game v3} \optag{jTeal}{NEW sign}}
\utt{8}{``Cancel just the welcome sign. Keep everything else.''}
  {\optag{jCoral}{CANCEL sign v1}}
\end{minipage}
}
\caption{A recorded five-minute session with \sysaudio. The user plans a game night while up to four tasks remain in flight. Each numbered utterance becomes one or more typed operations on the \ledger (tags below), with task-control operations bound to live task versions. Each bar is one version of a task. The game goes through four versions, the Python question gets a follow-up, a one-night budget of \$60 leaves the usual \$40 in memory untouched, and the cancellation removes the welcome sign and nothing else.}
\label{fig:teaser}
\end{figure}
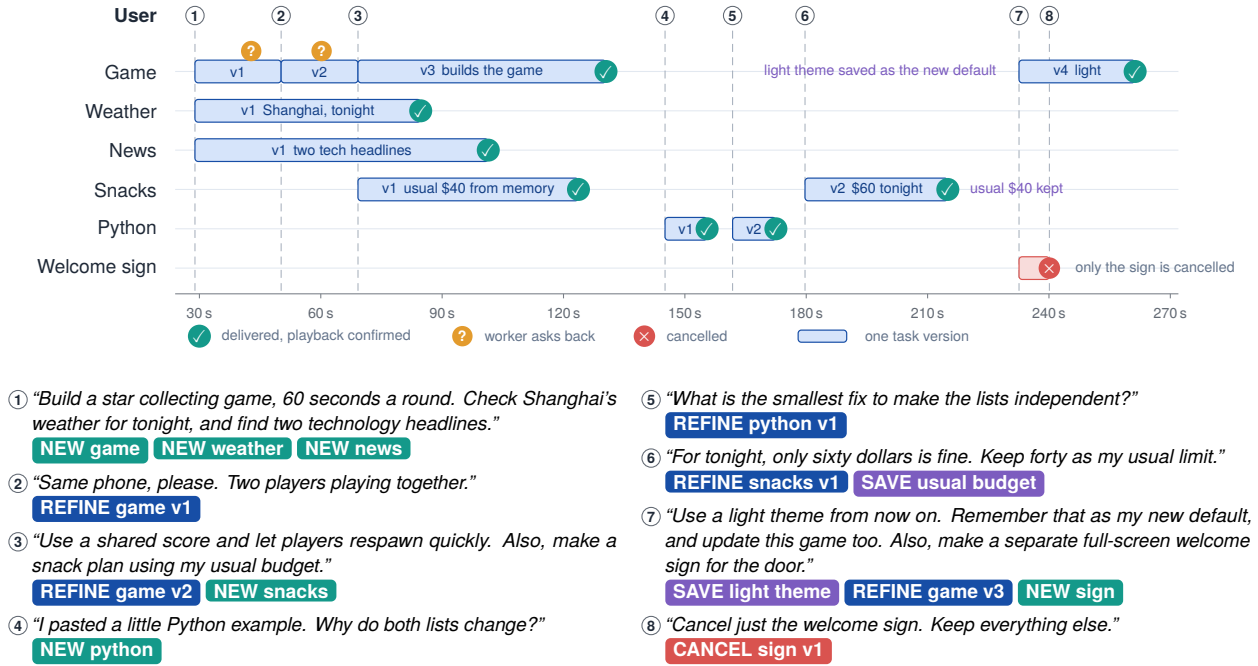

Figure~\ref{fig:teaser} shows what this means in practice. In a session recorded with \sysname, the harness we propose\footnote{The video and more sessions from daily use and games are at \url{https://amphionteam.github.io/Jarvis-Demo/}.}, the user asks in one breath for a browser game, tonight's weather in Shanghai and two technology headlines. Over four minutes the game is reshaped three times to add two players on one phone, a shared score with quick respawns and a light theme that also becomes the saved default. In between come a Python question with a follow-up, a sixty-dollar snack budget for tonight that leaves the usual forty intact, and a welcome sign that is later cancelled while everything else keeps running.

None of these utterances stands on its own. Each must be read against up to four tasks already in flight, so every time the system has to answer three questions. Does the utterance start new work or change a running task, and which one? Is a finished result still what the user wants? And did the user hear the whole answer, or was it cut off?

These are coordination decisions, yet current systems leave them to a component whose main job lies elsewhere. End-to-end models make them while they speak~\cite{gptrealtime2025,gemini_live_2026}, with only the text in their context to tell which task version is still live. Harnesses with a fast and a slow brain hand them to one of the two. The front model can emit task calls while it replies~\cite{venus2026,gander2026} but cannot see task versions. The backend agent can decide while reasoning about its task~\cite{qwen_audio_agent_2026} but does not follow the conversation. One hears every word and the other knows the task, but coordination needs both.

A task manager alone does not close this gap. Qwen Audio Agent (QAA)~\cite{qwen_audio_agent_2026} already tracks tasks, supports cancellation and collects playback receipts, yet a model still writes the call that maps an utterance to a task, and nothing checks that its target exists or is still current when it takes effect. ``Make it sixty dollars'' can then change the game instead of the snack plan, a correction can land on a replaced version, and a late draft can be played after the user has moved on. Table~\ref{tab:failures} lists six such failure modes, five with audio alone and a sixth once the agent can see.

These failures occur even when recognition is accurate and every worker reasons correctly about its request. The problem is timing. Because the two brains run at different speeds, a decision can rest on an incomplete view of the work or outlive the state that justified it. Better recognition and reasoning cannot keep them in step, and prompt rules such as ``follow the latest request'' still leave open which version is current when an action takes effect. We therefore argue that coordination deserves a brain of its own, one that sees both the conversation and the live tasks, backed by a runtime that enforces its decisions. Our guiding rule is \emph{models judge, code enforces}.

We propose \sysname, a coordination harness built on this rule. Every request becomes a task in the \ledger, a versioned record of what the user asked for and what happened to it, and every change becomes an operation on that ledger, such as refining version~3 of the game. A typed decision model, \jev\footnote{\url{https://typesafe.ai/}}, reads the new turn against the live tasks and picks among operations the runtime builds from the ledger, never writing one itself. Code checks the task and version of each pick before applying it, so no choice can name a missing task, and one aimed at a replaced version becomes a no-op.

The same rule runs through the rest of the harness. The \gate, trained on the front model's control tokens, decides when a turn goes to the slow brain without touching speech weights. If \jev is unsure, no task changes and a worker answers or asks the user to clarify. The \ledger accepts results only for live versions and rechecks plans against the conversation before they run, so stale work cannot take effect just because a worker finished. \delivery marks an answer delivered only after playback receipts show it played in full, and stopping the voice does not cancel the work behind it. \memory saves ``usually forty'' as a lasting fact and leaves ``only sixty tonight'' on the snack task.

We build two instances of this design. \sysaudio is built mainly on Realtime-Venus-Audio, and the same harness runs unchanged on MiniCPM-o~4.5 and NVIDIA's VoiceChat-11B~\cite{voicechat2026}. \sysomni pairs it with Venus-Omni and the multimodal \jevomni\footnote{\url{https://huggingface.co/akhilaaa3/Jev-Omni}}, reusing the ledger, workers and delivery layer. Vision brings a new source of outdated state, because a scene can change while nobody speaks. A teammate alive when the user asks for advice may be dead by the time the answer is ready, so \sysomni gives each visual fact a validity window and rechecks the evidence behind an answer before and during playback. An answer whose evidence expires is dropped or cut off.

Coordination alone, without a larger front model or a stronger worker, brings large gains (Figure~\ref{fig:headline}). With the front model, admission path and back end held fixed, replacing QAA with \sysname raises $\tau$-Voice~\cite{tauvoice2026} task success on Venus-Audio from 29 to 192 of 278 tasks, and the gains hold on every benchmark and front model. Against published systems, \sysaudio reaches 69\% Pass@1 on Full-Duplex-Bench v3~\cite{fullduplexbench_v3} with an open 9B front model, versus 60\% for GPT-Realtime, and leads all realtime speech models on the public $\tau$-Voice leaderboard (Table~\ref{tab:tau-published}).

\begin{table}[t]
  \centering
  \caption{Failure modes of a voice agent whose background work runs while the conversation continues, with the mechanism in \sysname that addresses each one and the evidence used to evaluate it.}
  \label{tab:failures}
  \tablefont
  \setlength{\tabcolsep}{5pt}
  \begin{tabularx}{\textwidth}{l >{\raggedright\arraybackslash}X >{\raggedright\arraybackslash}p{3.5cm} l}
  \toprule
  \tableheader \textbf{Failure mode} & \textbf{What the user experiences} & \textbf{Mitigated by} & \textbf{Evidence} \\
  \midrule
  Stale delivery & The first draft of a game is presented after the user has already asked for a different theme. & \ledger versions & Table~\ref{tab:stress} \\
  Misbinding & ``Make it sixty dollars'' changes the game instead of the snack plan. & \brain over version-bound options & Table~\ref{tab:abl-jev} \\
  Ghost execution & A plan formed before the user corrected the order number is executed anyway. & Freshness check on the plan & Table~\ref{tab:stress} \\
  Lost delivery & The user cuts an answer off halfway, yet the task is recorded as delivered. & Playback receipts and epochs & Table~\ref{tab:stress} \\
  Memory drift & ``Only sixty tonight'' overwrites the usual budget of forty. & Gated memory writes & Table~\ref{tab:memory} \\
  Expired evidence & Advice to push forward arrives after the teammate on screen has died. & \belief validity windows & \S\ref{sec:exp-omni} \\
  \bottomrule
  \end{tabularx}
  \end{table}

Our contributions are as follows.
\begin{itemize}[leftmargin=1.4em, itemsep=2pt]
  \item We identify coordination between a full-duplex front model and its background work as a problem in its own right, with six failure modes and four design principles (\S\ref{sec:overview}).
  \item We present \sysname, in which \jev selects runtime-built operations on a versioned \ledger and code checks each one before it takes effect, together with a native gate trained with GRPO, plan and result freshness checks, receipt-based delivery and gated memory (\S\ref{sec:audio}).
  \item We extend the design to audio-visual interaction with \sysomni, which treats visual evidence as time-bounded state and rechecks it before and during playback (\S\ref{sec:omni}).
  \item We show large and consistent gains from the harness alone across three benchmarks and three front models, and test each mechanism in controlled studies and live sessions (\S\ref{sec:experiments} and \S\ref{sec:applications}).
\end{itemize}

\section{Jarvis at a Glance}
\label{sec:overview}

Before describing the modules, we look at what goes wrong when nothing coordinates the two brains, and at the principles that follow.

\begin{figure}[t]
  \centering
\begin{tikzpicture}[font=\sffamily\footnotesize]
  \tikzset{
    mod/.style={rounded corners=3pt, draw=#1, fill=#1!10, line width=0.6pt, align=center,
                minimum height=10mm, inner sep=2.5pt, text=jInk},
    main/.style={minimum height=13.2mm},
    group/.style={rounded corners=5pt, draw=#1!70, dashed, line width=0.6pt, fill=#1!3},
    grouplab/.style={font=\sffamily\footnotesize\bfseries, text=#1, anchor=south, inner sep=2pt},
    flow/.style={-{Stealth[length=2mm]}, line width=0.7pt, draw=jInk!65},
    back/.style={-{Stealth[length=2mm]}, line width=0.7pt, draw=jCoral!80!black},
    lab/.style={font=\sffamily\scriptsize, text=jInk!80, inner sep=1.2pt},
  }

  \node[mod=jSlate, minimum width=8.6mm, inner sep=1.5pt, minimum height=29.6mm] (user) at (0.43, -0.82)
    {\textbf{User}\\[2pt]\scriptsize mic\\\scriptsize screen};
  \node[mod=jAmber, main, minimum width=23mm] (gate) at (3.11, 0)
    {\textbf{\gate}\\[1pt]\scriptsize LISTEN, DIRECT\\\scriptsize or DELEGATE};
  \node[mod=jSlate, minimum width=23mm] (talker) at (3.11, -1.8)
    {\textbf{Frozen speech}\\[1pt]\scriptsize Thinker and Talker};

  \node[mod=jSlate, main, minimum width=20mm] (bind) at (7.21, 0)
    {\textbf{Turn binding}\\[1pt]\scriptsize ASR observer};
  \node[mod=jBlue, main, minimum width=22mm] (jev) at (9.86, 0)
    {\textbf{\brain}\\[1pt]\scriptsize \jev};
  \node[mod=jTeal, main, minimum width=19mm] (ledger) at (13.24, 0)
    {\textbf{\ledger}\\[1pt]\scriptsize versions\\\scriptsize freshness};
  \node[mod=jTeal, main, minimum width=14.5mm] (workers) at (15.615, 0)
    {\textbf{Workers}\\[1pt]\scriptsize async\\\scriptsize tool loops};
  \node[mod=jSlate, dashed, minimum width=30mm, anchor=west] (evid) at (bind.west |- 0, 1.8)
    {\textbf{\timeline}\\[1pt]\scriptsize \belief (Omni)};
  \node[mod=jPlum, minimum width=25mm] (mem) at (13.24, 1.8)
    {\textbf{\memory}\\[1pt]\scriptsize profile \textbar\ task records};
  \node[mod=jCoral, minimum width=21mm] (deliv) at (13.24, -1.8)
    {\textbf{\delivery}\\[1pt]\scriptsize epochs \textbar\ receipts};
  \node[mod=jSlate, minimum width=15mm] (tools) at (15.615, -1.95)
    {\textbf{Tools}\\[1pt]\scriptsize web \textbar\ files\\\scriptsize games};

  \begin{scope}[on background layer]
    \node[group=jAmber, fit=(gate)(talker), inner sep=4pt] (fast) {};
    \node[grouplab=jAmber!80!black] at (fast.north) {Fast brain};
    \path[group=jBlue]
      ($(bind.west |- evid.north)+(-6pt,6pt)$) coordinate (slownw)
      -- ($(workers.east |- evid.north)+(6pt,6pt)$) coordinate (slowne)
      -- ($(workers.south east)+(6pt,-6pt)$)
      -- ($(deliv.east |- workers.south)+(6pt,-6pt)$)
      -- ($(deliv.south east)+(6pt,-6pt)$)
      -- ($(bind.west |- deliv.south)+(-6pt,-6pt)$) -- cycle;
    \node[grouplab=jBlue] at ($(slownw)!0.5!(slowne)$) {Slow brain (\sysname harness)};
  \end{scope}

  \draw[flow] (user.east |- gate.west) -- (gate.west) node[lab, midway, above] {audio};
  \draw[flow] (gate.east) -- (bind.west) node[lab, midway, above] {DELEGATE};
  \draw[flow] (gate.south) -- (talker.north) node[lab, midway, right] {DIRECT};
  \draw[flow] (bind.east) -- (jev.west);
  \draw[flow, dashed] (evid.east) -| (jev.north) node[lab, pos=0.8, left] {visual claims};
  \draw[flow, <->] (mem.west) -| (jev.north east);
  \draw[flow, <->] (mem.south) -- (ledger.north);
  \draw[flow] (jev.east) -- (ledger.west) node[lab, midway, above] {typed ops};
  \draw[flow, <->] (ledger.east) -- (workers.west);
  \draw[flow, <->] (workers.south) -- (workers.south |- tools.north);
  \draw[flow] (ledger.south) -- (deliv.north) node[lab, midway, left] {live results};
  \draw[back] ($(deliv.west)+(0,0.13)$) -- ($(talker.east)+(0,0.13)$)
    node[lab, pos=0.42, above] {speech request over the backend channel};
  \draw[back] ($(talker.east)+(0,-0.13)$) -- ($(deliv.west)+(0,-0.13)$)
    node[lab, midway, below] {playback receipts};
  \draw[flow] (talker.west) -- (talker.west -| user.east) node[lab, midway, above] {speech};
\end{tikzpicture}
  \caption{Overview of \sysname. The fast brain keeps the conversation going and decides when to hand a turn over. The slow brain interprets the turn with \jev, records the resulting work in the \ledger, runs it asynchronously and returns the answer through the fast brain's own voice once the result is still valid. The dashed module is used by \sysomni only.}
  \label{fig:overview}
  \end{figure}
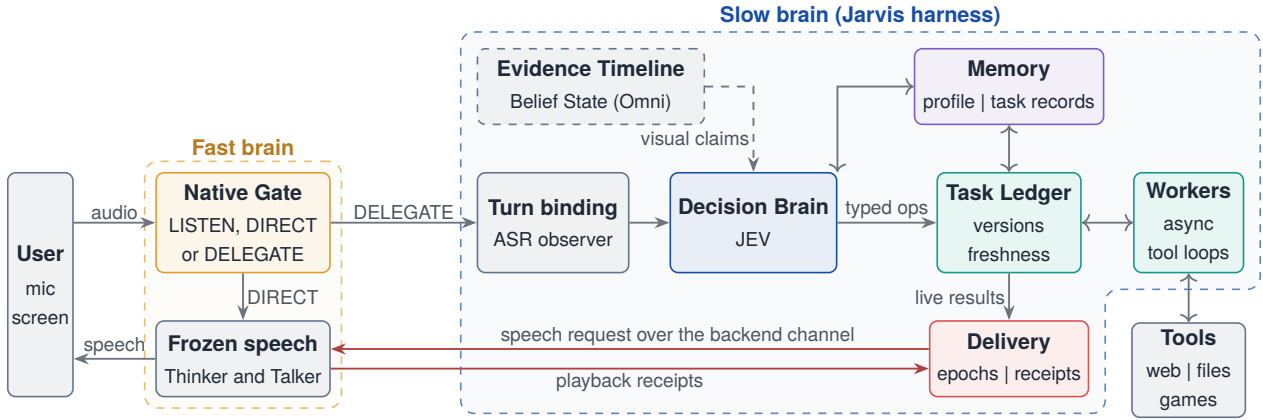

\subsection{What Goes Wrong Without Coordination}
\label{sec:failures}

Running background work under a live conversation produces its own class of errors. They are not recognition errors and not reasoning errors. They happen because two processes move at different speeds and nothing keeps them in step. Table~\ref{tab:failures} lists the six failure modes we see. The first five arise with audio alone, and the sixth appears once the agent can see.

All six share one cause. The system acts on a picture of the conversation that has stopped being true, whether that picture is a task version, a plan, a playback state, a remembered fact or a frame of video. Patching each case with prompt rules does not scale, because every new situation produces a new variant. \sysname removes the cause instead. It makes the current state explicit and checks every action against it.

\subsection{Design Principles}
\label{sec:principles}

Four principles follow, and every module of \sysname applies at least one of them.

\begin{enumerate}[leftmargin=1.4em, itemsep=3pt]
  \item \textbf{Learn \emph{when}, keep \emph{how}.} The front model already speaks well. We train only the decision of when to involve the slow brain, on the model's own control tokens, and leave every weight that produces speech frozen.
  \item \textbf{Models judge, code enforces.} Questions that need understanding go to a decision model, but the model answers by choosing among operations that the runtime has prepared. The runtime then applies the chosen operation and checks it against the current state.
  \item \textbf{Nothing is assumed current.} Plans, results, playback and visual facts all carry a version, an epoch or a timestamp, and each is checked again before it may act or speak.
  \item \textbf{Done is not delivered.} A worker finishing a task and the user hearing the answer are two separate events, and the system records them separately.
\end{enumerate}

\subsection{How a Request Flows}
\label{sec:flow}

Figure~\ref{fig:overview} shows the path of a request. Audio reaches the front model and a streaming ASR observer at the same time. For every one-second audio unit, the \gate reads the control state of the front model and chooses to keep listening, to let the front model answer on its own, or to hand the turn to the slow brain. Small talk and short answers therefore stay fast. A hand-off carries the audio sample at which it was made, so the slow brain works from the transcript of what the user actually said rather than from text the front model generated.

\jev selects operations, the \ledger applies them and runs workers, and \delivery speaks accepted results in the front model's own voice. The two brains never wait for each other. The fast brain keeps talking while three workers run, and the slow brain updates its tasks while the user is still speaking.

This design has two instances. \sysaudio, the main subject of this report, runs on full-duplex speech models. \sysomni runs on Venus-Omni, which also watches the screen. It keeps the \ledger, the workers and \delivery unchanged, replaces \jev with the multimodal \jevomni, and adds the dashed perception layer of Figure~\ref{fig:overview}, which keeps visual facts together with the time they stop being valid. Section~\ref{sec:audio} describes the modules in \sysaudio, Section~\ref{sec:omni} shows what changes when the front model can see, and Table~\ref{tab:instances} in Appendix~\ref{app:impl} compares the two instances side by side.

\section{The Jarvis Harness}
\label{sec:audio}

This section follows the game night of Figure~\ref{fig:teaser} through the five modules of \sysname, as they run in \sysaudio on Realtime-Venus-Audio~\cite{venus2026}. Venus-Audio is a 9B full-duplex model that follows the Omni-Flow design of MiniCPM-o~4.5. A Whisper-Medium encoder feeds a Qwen3-8B Thinker, whose text and hidden states drive a streaming Talker. The model consumes audio in one-second units, predicts at every unit whether to listen or to speak, and has a native \texttt{<delegate>} token for handing work to a backend. We use the released checkpoint as the fast brain, and apart from a small control adapter none of its weights change.

\subsection{Native Gate: Learning When to Hand Off}
\label{sec:gate}

The first decision in every turn is whether the slow brain should be involved at all. ``Build me a little game for tonight'' needs it, ``tell me a joke'' does not, and ``hold on, let me ask my roommate'' is not addressed to the assistant. The decision has to be made from audio while the user may still be talking, and a rule on top of ASR cannot make it well, because the words alone do not tell a finished request from a pause. The front model already carries the information that matters, since it is the part of the system that decides when to speak. The \gate therefore reads its decision directly from the front model's native control tokens.

\paragraph{A policy over native tokens.}
Let $x$ be the causal prefix of the conversation up to the current audio unit. Venus-Audio first predicts \texttt{<|listen|>} or \texttt{<|speak|>}, and after \texttt{<|speak|>} it either emits \texttt{<delegate>} or starts a public reply. We read a three-way policy off these two positions,
\begin{equation}
\begin{aligned}
\pi(\act{LISTEN}\mid x) &= P(\texttt{<|listen|>}\mid x),\\
\pi(\act{DIRECT}\mid x) &= P(\texttt{<|speak|>}\mid x)\,\textstyle\sum_{w\in\mathcal{V}_{\mathrm{pub}}} P(w\mid x,\texttt{<|speak|>}),\\
\pi(\act{DELEGATE}\mid x) &= P(\texttt{<|speak|>}\mid x)\,P(\texttt{<delegate>}\mid x,\texttt{<|speak|>}),
\end{aligned}
\label{eq:policy}
\end{equation}
where $\mathcal{V}_{\mathrm{pub}}$ is the set of ordinary tokens that may begin a spoken reply. The \act{DIRECT} branch sums over every possible first word, so the policy decides \emph{whether} to answer directly without committing to \emph{what} the answer starts with. A \act{DIRECT} or \act{DELEGATE} action fires only when its probability reaches 0.8, and every other case falls back to \act{LISTEN}. When the gate is unsure, it keeps listening.

\begin{figure}[t]
\centering
\begin{tikzpicture}[font=\sffamily\scriptsize]
  \begin{scope}[x=0.82cm, y=2.55cm]
    \fill[jSlateSoft!60] (0,0) rectangle (1.5,1);
    \fill[jAmberSoft!70] (1.5,0) rectangle (5.0,1);
    \fill[jTealSoft!60] (5.0,0) rectangle (8,1);
    \node[font=\sffamily\tiny, text=jSlate] at (0.75,1.07) {silence};
    \node[font=\sffamily\tiny, text=jAmber!70!black] at (3.25,1.07) {user speaking};
    \node[font=\sffamily\tiny, text=jTeal!70!black] at (6.5,1.07) {request complete};
    \node[font=\sffamily\tiny\itshape, text=jInk!75, anchor=north] at (4,-0.13) {``Check Shanghai's weather for tonight.''};

    \draw[jInk!60, line width=0.4pt] (0,0) -- (8,0);
    \draw[jInk!60, line width=0.4pt] (0,0) -- (0,1);
    \foreach \u in {1,...,8} {
      \draw[jInk!40] (\u,0) -- (\u,-0.025);
    }
    \foreach \p in {0.5,1} \node[font=\sffamily\tiny, text=jInk!60, left] at (0,\p) {\p};
    \draw[jCoral, densely dashed, line width=0.5pt] (0,0.8) -- (8,0.8);
    \node[font=\sffamily\tiny, text=jCoral, anchor=north west] at (2.0,0.79) {threshold 0.8};
    \node[font=\sffamily\tiny, text=jInk!60, anchor=north east] at (8,-0.13) {1\,s units};

    \draw[jSlate, line width=1pt] plot[smooth, tension=0.6] coordinates
      {(0.5,0.97) (1.5,0.96) (2.5,0.93) (3.5,0.9) (4.5,0.86) (5.3,0.15) (6.5,0.05) (7.5,0.04)};
    \draw[jTeal, line width=1pt] plot[smooth, tension=0.6] coordinates
      {(0.5,0.02) (1.5,0.03) (2.5,0.05) (3.5,0.07) (4.5,0.08) (5.3,0.06) (6.5,0.05) (7.5,0.05)};
    \draw[jBlue, line width=1.1pt] plot[smooth, tension=0.6] coordinates
      {(0.5,0.01) (1.5,0.01) (2.5,0.02) (3.5,0.03) (4.5,0.06) (5.3,0.79) (6.5,0.9) (7.5,0.91)};
    \fill[jBlue] (5.33,0.81) circle (1.6pt);
    \node[font=\sffamily\tiny, text=jBlue, anchor=west, align=left] at (5.5,0.55) {DELEGATE fires\\0.16\,s after the end\\(median on dev)};

    \node[font=\sffamily\tiny, text=jSlate, anchor=north west, fill=white, fill opacity=0.85, text opacity=1, inner sep=1pt] at (0.15,0.94) {$\pi(\textsf{LISTEN})$};
    \node[font=\sffamily\tiny, text=jTeal, anchor=west] at (1.6,0.14) {$\pi(\textsf{DIRECT})$};
    \node[font=\sffamily\tiny, text=jBlue, anchor=south west] at (6.0,0.91) {$\pi(\textsf{DELEGATE})$};
    \node[font=\sffamily\scriptsize\bfseries, anchor=north west] at (-0.6,1.32) {(a)};
  \end{scope}

  \begin{scope}[shift={(7.6,0)}]
    \tikzset{
      blk/.style={rounded corners=2.5pt, draw=#1, fill=#1!10, line width=0.55pt, align=center,
                  minimum height=7mm, inner sep=2.5pt, text=jInk},
      ar/.style={-{Stealth[length=1.7mm]}, line width=0.55pt, draw=jInk!65},
    }
    \node[font=\sffamily\scriptsize\bfseries, anchor=north west] at (-0.35,3.37) {(b)};
    \node[blk=jSlate, minimum width=13mm] (x) at (0.45,1.3) {causal\\prefix $x$};

    \node[blk=jAmber, minimum width=24mm] (route) at (3.0,2.3) {Thinker + LoRA\\\tiny q, v, o \quad rank 8};
    \node[blk=jBlue, minimum width=19mm] (pi) at (5.75,2.3) {$\pi(\cdot\mid x)$\\\tiny 3 branches};
    \node[blk=jSlate, minimum width=24mm] (speech) at (3.0,0.3) {Thinker, frozen\\\tiny adapter off};
    \node[blk=jSlate, minimum width=19mm] (talk) at (5.75,0.3) {Talker\\\tiny speech out};

    \draw[ar] (x.north) |- (route.west) node[font=\sffamily\tiny, text=jInk!75, pos=0.75, above] {routing pass};
    \draw[ar] (x.south) |- (speech.west) node[font=\sffamily\tiny, text=jInk!75, pos=0.75, below] {speech pass};
    \draw[ar] (route.east) -- (pi.west);
    \draw[ar] (speech.east) -- (talk.west);

    \node[font=\sffamily\tiny, text=jInk!80, align=left, anchor=north] at (4.35,1.85)
      {position 1: \texttt{<|listen|>} vs \texttt{<|speak|>}\\position 2: \texttt{<delegate>} vs public tokens};
    \draw[jSlate!70, densely dotted, line width=0.5pt] (1.55,1.2) -- (7.0,1.2);
    \node[font=\sffamily\tiny, text=jSlate, anchor=west] at (1.55,1.04) {separate key-value caches};
  \end{scope}
\end{tikzpicture}
\caption{The \gate. (a) On each one-second audio unit the gate reads a three-way policy from the native tokens of the front model. It keeps listening through leading silence and through an unfinished request, and fires \act{DELEGATE} once the request is complete and the probability passes 0.8. (b) The LoRA adapter is active only in the routing pass. Speech is produced by the frozen weights with a separate key-value cache.}
\label{fig:gate}
\end{figure}
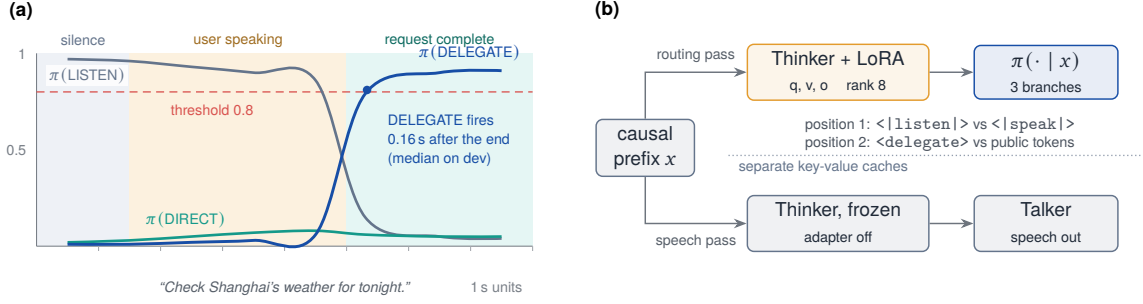

\paragraph{Training only the decision.}
The adapter is a LoRA~\cite{hu2022lora} of rank 8 on the \texttt{q}, \texttt{v} and \texttt{o} projections of the Thinker, 6.2M parameters or about 0.07\% of the model. It is switched on only inside the forward pass that computes Equation~(\ref{eq:policy}), and that pass builds its own prefix instead of sharing the key-value cache of the speech pass (Figure~\ref{fig:gate}b). The voice the user hears is therefore produced by exactly the released weights.

We train the adapter with group-relative policy optimization (GRPO)~\cite{shao2024deepseekmath} on speech we authored and synthesized ourselves. We wrote 264 bilingual scenarios, from small talk and tool lookups to task corrections and remarks meant for someone else, rendered them with Qwen3-TTS~\cite{qwen3tts2026} and encoded them with the streaming encoder of Venus-Audio. Each utterance is cut into causal prefixes. A prefix that ends before the request is complete is labeled \act{LISTEN}, so a single recording teaches both what to do and when it is safe to do it. For each prefix the gate samples eight actions, a correct action earns $+2$ and a wrong one $-1$, and a KL penalty keeps the policy close to where each stage started. Because the reward is on the branch and not on a token, the optimizer raises the chance of answering directly without teaching the model how to begin its answers. A four-stage curriculum adds telephone speech and conversational requests such as asking for a story. On held-out scenario families, the final gate makes 98.8\% of first admissions at the right time and in the right branch (162 of 164), and every case from earlier stages stays correct. Appendix~\ref{app:gate} gives the data, the objective and the curriculum.

\paragraph{Binding the turn.}
The gate decides when, and a streaming ASR observer, Qwen3-ASR-1.7B~\cite{qwen3asr2026}, supplies the words. When the gate fires \act{DELEGATE}, the event records the audio sample at which the decision was made, and the runtime attaches the most recent final transcript that started before that sample. The slow brain thus receives what the user actually said, including names, numbers and spelled identifiers, rather than a paraphrase generated by the front model.

\subsection{Decision Brain: Typed Judgments with JEV}
\label{sec:brain}

Once a turn is admitted, \sysname has to work out what it means for the work in flight. Midway through the game night the user says ``make it sixty dollars''. At that moment the game is on its third version, the snack plan has already been delivered and the answer to the Python follow-up has just been played. Understanding the sentence needs a model. Acting on it changes real tasks, so the model must never produce an operation that the runtime cannot check. \jev meets both needs because it is a typed judgment model rather than a text generator.

\paragraph{The typed interface.}
A request to \jev consists of a state and a set of typed questions. The state holds the bound transcript, the last twelve turns, any question the assistant is waiting on, the user profile and the available tools. Each question belongs to one of three primitives. A \emph{Choice} question lists named options and returns a probability distribution over them. A \emph{Noul} question returns the probability of yes for a single statement. A \emph{Score} question returns an expected value on an ordinal scale. None of them returns free text, so every answer is an option the runtime wrote or a number.

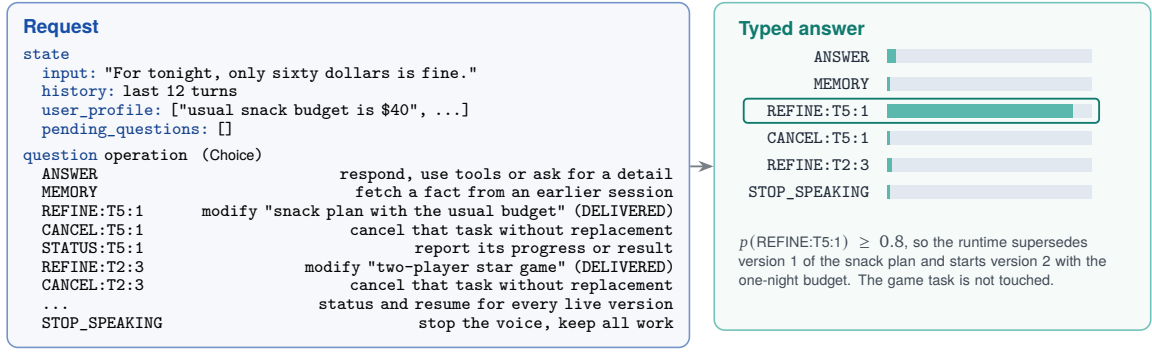
\begin{figure}[t]
\centering
\begin{tikzpicture}[font=\sffamily\scriptsize]
  \tikzset{
    panel/.style={rounded corners=4pt, draw=#1!60, fill=#1!4, line width=0.6pt, inner sep=6pt},
    key/.style={font=\ttfamily\tiny, text=jBlue!85!black},
    val/.style={font=\ttfamily\tiny, text=jInk},
  }
  \node[panel=jBlue, anchor=north west, text width=8.6cm] (req) at (0,0) {%
    {\sffamily\scriptsize\bfseries\color{jBlue} Request}\par\vspace{3pt}
    {\ttfamily\tiny
    \textcolor{jBlue!85!black}{state}\par
    \hspace*{1em}\textcolor{jBlue!85!black}{input:}\ ``For tonight, only sixty dollars is fine.''\par
    \hspace*{1em}\textcolor{jBlue!85!black}{history:}\ last 12 turns\par
    \hspace*{1em}\textcolor{jBlue!85!black}{user\_profile:}\ [``usual snack budget is \$40'', \ldots]\par
    \hspace*{1em}\textcolor{jBlue!85!black}{pending\_questions:}\ []\par
    \vspace{3pt}
    \textcolor{jBlue!85!black}{question}\ operation \ (\textsf{Choice})\par
    \hspace*{1em}ANSWER\hfill respond, use tools or ask for a detail\par
    \hspace*{1em}MEMORY\hfill fetch a fact from an earlier session\par
    \hspace*{1em}REFINE:T5:1\hfill modify ``snack plan with the usual budget'' (DELIVERED)\par
    \hspace*{1em}CANCEL:T5:1\hfill cancel that task without replacement\par
    \hspace*{1em}STATUS:T5:1\hfill report its progress or result\par
    \hspace*{1em}REFINE:T2:3\hfill modify ``two-player star game'' (DELIVERED)\par
    \hspace*{1em}CANCEL:T2:3\hfill cancel that task without replacement\par
    \hspace*{1em}\ldots\hfill status and resume for every live version\par
    \hspace*{1em}STOP\_SPEAKING\hfill stop the voice, keep all work\par
    }};

  \node[panel=jTeal, anchor=north west, text width=5.35cm, minimum height=4.33cm] (ans) at (9.35,0) {};
  \node[font=\sffamily\scriptsize\bfseries, text=jTeal!80!black, anchor=north west] at (9.55,-0.12) {Typed answer};
  \foreach \lab/\p/\i in {ANSWER/0.04/0, MEMORY/0.01/1, REFINE:T5:1/0.91/2, CANCEL:T5:1/0.01/3, REFINE:T2:3/0.02/4, STOP\_SPEAKING/0.01/5} {
    \pgfmathsetmacro{\yy}{-0.72-0.36*\i}
    \node[font=\ttfamily\tiny, text=jInk, anchor=east] at (11.55,\yy) {\lab};
    \fill[jSlateSoft] (11.65,\yy-0.09) rectangle (14.35,\yy+0.09);
    \fill[jTeal!70] (11.65,\yy-0.09) rectangle ({11.65+2.7*\p},\yy+0.09);
  }
  \draw[jTeal!80!black, line width=0.7pt, rounded corners=1.5pt] (9.75,-1.6) rectangle (14.45,-1.28);
  \node[font=\sffamily\tiny, text=jInk!85, anchor=north west, text width=4.9cm] at (9.55,-2.95)
    {$p(\text{REFINE:T5:1}) \ge 0.8$, so the runtime supersedes version 1 of the snack plan and starts version 2 with the one-night budget. The game task is not touched.};
  \draw[-{Stealth[length=2mm]}, line width=0.7pt, draw=jInk!60] (req.east |- ans.west) -- (ans.west);
\end{tikzpicture}
\caption{A decision request to \jev, simplified from the runtime payload. The runtime writes the options from the live \ledger, so the answer is a distribution over operations that the runtime can execute.}
\label{fig:jev}
\end{figure}

\paragraph{Options come from the ledger.}
The key design choice is who writes the options. For an admitted turn, the runtime builds them from the current state of the \ledger,
\begin{equation}
\mathcal{O}_t = \{\op{ANSWER}, \op{MEMORY}\} \;\cup\; \{\,a{:}\tau{:}v \;\mid\; a\in\{\op{CANCEL},\op{REFINE},\op{STATUS},\op{RESUME}\},\ (\tau,v)\in\mathcal{L}_t\,\} \;\cup\; \mathcal{S}_t,
\label{eq:options}
\end{equation}
where $\mathcal{L}_t$ contains the live versions of the most recent tasks and $\mathcal{S}_t$ adds \op{STOP\_SPEAKING} while audio is playing. Each task option carries the objective and status of its task in plain words. For ``make it sixty dollars'' the options include \op{REFINE:snacks:1} and \op{REFINE:game:3}, and \jev returns a probability for each. The runtime applies
\begin{equation}
o^{\ast}=\arg\max_{o\in\mathcal{O}_t} p(o\mid s_t), \qquad \text{execute } o^{\ast} \text{ only if } p(o^{\ast}\mid s_t)\ge\theta .
\label{eq:accept}
\end{equation}
Because every option names a task and a version, an accepted answer is executable by construction. \jev cannot refer to a task that does not exist, and a choice made against version~2 becomes a no-op if version~3 has appeared in the meantime. When the confidence is below $\theta=0.8$, the runtime does not guess. It passes the turn to a worker with full context, which either answers or asks the user one specific question, and every existing task stays as it was. Figure~\ref{fig:jev} shows a request of this kind.

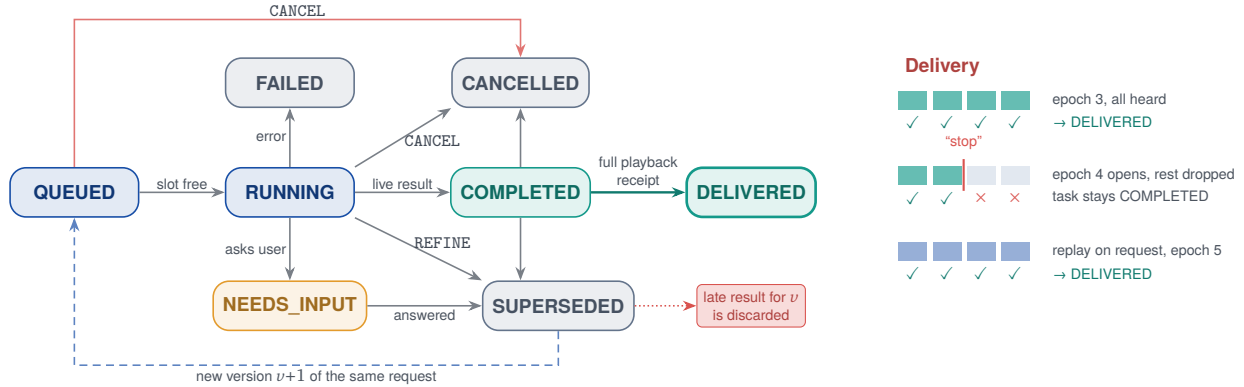
\begin{figure}[t]
  \centering
\begin{tikzpicture}[font=\sffamily\scriptsize]
  \tikzset{
    st/.style={rounded corners=6pt, draw=#1, fill=#1!12, line width=0.6pt, minimum width=17mm,
               minimum height=6.5mm, font=\sffamily\scriptsize\bfseries, text=#1!70!black},
    tr/.style={-{Stealth[length=1.8mm]}, line width=0.6pt, draw=jInk!60},
    lb/.style={font=\sffamily\tiny, text=jInk!80, inner sep=1pt},
    opl/.style={font=\ttfamily\tiny, text=jInk!85, inner sep=1pt},
  }
  \node[st=jBlue] (q) at (0,0) {QUEUED};
  \node[st=jBlue] (r) at (2.85,0) {RUNNING};
  \node[st=jTeal] (c) at (5.9,0) {COMPLETED};
  \node[st=jTeal, line width=1pt] (d) at (8.95,0) {DELIVERED};
  \node[st=jAmber] (n) at (2.85,-1.5) {NEEDS\_INPUT};
  \node[st=jSlate] (f) at (2.85,1.45) {FAILED};
  \node[st=jSlate] (x) at (5.9,1.45) {CANCELLED};
  \node[st=jSlate] (s) at (6.4,-1.5) {SUPERSEDED};

  \draw[tr] (q) -- (r) node[lb, midway, above] {slot free};
  \draw[tr] (r) -- (c) node[lb, midway, above] {live result};
  \draw[tr, draw=jTeal!80!black, line width=0.9pt] (c) -- (d) node[lb, midway, above, align=center] {full playback\\receipt};
  \draw[tr] (r) -- (n) node[lb, midway, left] {asks user};
  \draw[tr] (r) -- (f) node[lb, midway, left] {error};
  \draw[tr] (r.north east) -- (x.south west) node[opl, pos=0.45, right=1pt] {CANCEL};
  \draw[tr] (c.north) -- (x.south);
  \draw[tr] (r.south east) -- (s.north west) node[opl, pos=0.4, right=2pt] {REFINE};
  \draw[tr] (n.east) -- (s.west) node[lb, midway, below] {answered};
  \draw[tr] (c.south) -- (c.south |- s.north);
  \draw[tr, densely dashed, draw=jBlue!70] (s.south) -- ++(0,-0.45) -| (q.south) node[lb, pos=0.25, below] {new version $v{+}1$ of the same request};
  \draw[tr, draw=jCoral!80] (q.north) -- ++(0,1.95) -| (x.north) node[opl, pos=0.25, above] {CANCEL};

  \node[draw=jCoral, fill=jCoralSoft, rounded corners=2pt, font=\sffamily\tiny, text=jCoral!80!black, align=center, inner sep=2.5pt] (late) at (8.95,-1.5) {late result for $v$\\is discarded};
  \draw[tr, draw=jCoral, densely dotted] (s.east) -- (late.west);

  \begin{scope}[shift={(10.9,1.1)}]
    \node[font=\sffamily\scriptsize\bfseries, text=jCoral!80!black, anchor=west] at (-0.05,0.55) {Delivery};
    \foreach \i in {0,...,3} {
      \fill[jTeal!65] (0.45*\i,0) rectangle (0.45*\i+0.38,0.24);
      \node[font=\sffamily\tiny, text=jTeal!80!black] at (0.45*\i+0.19,-0.17) {\checkmark};
    }
    \node[font=\sffamily\tiny, text=jInk!80, anchor=west] at (1.9,0.12) {epoch 3, all heard};
    \node[font=\sffamily\tiny, text=jTeal!80!black, anchor=west] at (1.9,-0.17) {\textrightarrow\ DELIVERED};

    \foreach \i in {0,1} {
      \fill[jTeal!65] (0.45*\i,-1.0) rectangle (0.45*\i+0.38,-0.76);
      \node[font=\sffamily\tiny, text=jTeal!80!black] at (0.45*\i+0.19,-1.17) {\checkmark};
    }
    \foreach \i in {2,3} {
      \fill[jSlateSoft] (0.45*\i,-1.0) rectangle (0.45*\i+0.38,-0.76);
      \node[font=\sffamily\tiny, text=jCoral] at (0.45*\i+0.19,-1.17) {$\times$};
    }
    \draw[jCoral, line width=0.8pt] (0.86,-1.08) -- (0.86,-0.66);
    \node[font=\sffamily\tiny, text=jCoral, anchor=south] at (0.86,-0.66) {``stop''};
    \node[font=\sffamily\tiny, text=jInk!80, anchor=west] at (1.9,-0.88) {epoch 4 opens, rest dropped};
    \node[font=\sffamily\tiny, text=jInk!80, anchor=west] at (1.9,-1.17) {task stays COMPLETED};

    \foreach \i in {0,...,3} {
      \fill[jBlue!45] (0.45*\i,-2.0) rectangle (0.45*\i+0.38,-1.76);
      \node[font=\sffamily\tiny, text=jTeal!80!black] at (0.45*\i+0.19,-2.17) {\checkmark};
    }
    \node[font=\sffamily\tiny, text=jInk!80, anchor=west] at (1.9,-1.88) {replay on request, epoch 5};
    \node[font=\sffamily\tiny, text=jTeal!80!black, anchor=west] at (1.9,-2.17) {\textrightarrow\ DELIVERED};
  \end{scope}
\end{tikzpicture}
  \caption{Life of a task in the \ledger. A refinement supersedes the live version and starts a new one, and a result that arrives for a superseded or cancelled version is discarded. \state{COMPLETED} becomes \state{DELIVERED} only when a playback receipt shows that the whole answer was heard in the current generation epoch. Stopping the voice opens a new epoch but leaves the task itself untouched.}
  \label{fig:ledger}
  \end{figure}

\paragraph{One brain for many decisions.}
The same interface serves every decision in the harness that needs understanding, from routing a turn and choosing its target task to gating memory writes and deciding when to deliver. Table~\ref{tab:jev-decisions} in the appendix lists all of them. Several small questions about one turn go to \jev in a single call and are answered independently. A decision on an admitted turn takes 0.81~s at the median and the memory-write gate 0.32~s. Speed matters as much as accuracy here, because a cancellation is useful only if it stops the old worker before that worker acts, and Section~\ref{sec:exp-jev} shows that this is exactly where general LLM controllers fall short.

Typed options also remove a confusion that free-form routing makes easy. ``Cancel my order'' in a retail conversation is a domain action that the worker must carry out with the store's tools, so it is an \op{ANSWER}. ``Cancel the web page you were writing'' stops one of the assistant's own jobs, so it is a \op{CANCEL} on that task. The same holds for short replies. A name, a number or a ``yes'' that answers a pending question continues the current service as an \op{ANSWER} rather than being dropped for lack of a new goal. When one utterance holds several requests, as in the first turn of Figure~\ref{fig:teaser}, a planner splits it into verbatim spans and each span is interpreted on its own, under the same checks as any other choice (Appendix~\ref{app:impl}).

\subsection{Task Ledger and Workers}
\label{sec:ledger}

The \ledger is the single place where \sysname records what the user has asked for and what has happened to it. Each task has an identifier, a version and a logical request identifier derived from the utterance that created it. Tasks created by the same utterance share a request group and keep their order, so the three requests in the first turn of Figure~\ref{fig:teaser} remain three tasks that report back in a sensible sequence. A task moves through the eight states of Figure~\ref{fig:ledger}.

\paragraph{Invariants.}
The ledger enforces four invariants, and together they rule out the first three failure modes of Table~\ref{tab:failures}.
\begin{enumerate}[leftmargin=1.6em, itemsep=2pt, label=\textbf{I\arabic*.}]
  \item \emph{One live version.} When the user asks for a shared score, version~2 of the game becomes version~3 and version~2 moves to \state{SUPERSEDED}. At most one version of a request is live at any time, and a repeated utterance creates no duplicate.
  \item \emph{Late results are rejected.} A worker result is accepted only if its task version is live. A slow draft of version~2 that finishes after the user asked for version~3 is discarded and logged, and it can never be presented.
  \item \emph{Plans are checked against the conversation.} A worker records the conversation snapshot it planned from, and before it executes a planned action the runtime compares that snapshot with the current one. If the user has corrected an order number or withdrawn an authorization in the meantime, the old plan does not run and the worker plans again.
  \item \emph{Delivery needs a receipt.} A task becomes \state{DELIVERED} only through a complete playback receipt (Section~\ref{sec:delivery}).
\end{enumerate}
Section~\ref{sec:exp-race} removes the checks behind I2 to I4 one at a time and measures what each prevents.

\paragraph{Workers.}
Each task runs in an asynchronous tool loop built on DeepSeek-V4-Flash~\cite{xu2026deepseek}, with up to three workers at once. Tool results are treated as evidence and never as new instructions, a repeated identical call returns the earlier receipt instead of running again, and a worker asks the user for a missing detail instead of guessing. A worker often produces files, such as the page of the browser game. The runtime builds the list of these files only from writes that the file tool has confirmed, never from what the model says it has written, so the assistant cannot tell the user about a file that does not exist.

\subsection{Delivery: From Done to Heard}
\label{sec:delivery}

A finished result is not yet an answer the user has received. Suppose the weather report starts playing and the user cuts in after three words to change the game. If the system marks the weather as delivered, the user never hears it and the assistant believes it has answered. \delivery keeps these two events apart with two mechanisms.

\paragraph{Generation epochs.}
Every time playback is cleared, for example because the user started speaking or said ``stop'', the runtime opens a new generation epoch. Every audio chunk and control message carries the epoch in which it was produced, and messages from an older epoch are dropped on arrival. Audio that was already in flight therefore cannot resume after the user has moved on.

\paragraph{Playback receipts.}
The client reports what it has actually played as a stream of receipts, each naming the epoch, the chunk, the number of samples played and whether playback was interrupted. For a presentation $r$ of task version $(\tau,v)$,
\begin{equation}
\textsc{Delivered}(\tau,v) \iff \exists\, r:\ e(r)=e_{\mathrm{now}} \ \wedge\ \textstyle\sum_{c\in r}\mathrm{played}(c)=\sum_{c\in r}\mathrm{samples}(c) \ \wedge\ \neg\,\mathrm{interrupted}(r).
\label{eq:delivered}
\end{equation}
Only this condition moves a task from \state{COMPLETED} to \state{DELIVERED}. The interrupted weather report is recorded as partially heard and keeps its result, so the user can ask for it again. Spoken results are voiced by the front model's own Talker through its private backend channel, so the assistant sounds the same whether it answers directly or reports on background work.

\paragraph{Stopping is not cancelling.}
``Stop talking'' and ``cancel that'' mean different things, and \sysname keeps them apart. Stopping the voice opens a new epoch and holds pending announcements, while every task keeps running and keeps its result. Cancelling invalidates one task version and stops its worker, and nothing else is affected.

\subsection{Memory: Durable Facts Without Drift}
\label{sec:memory}

A voice assistant that people use every day should remember what they tell it, but not everything they say. ``I usually spend forty dollars'' is a fact worth keeping. ``Only sixty tonight'' is a condition on tonight's task, and ``tell him I'll call back'' was not said to the assistant at all.

\memory keeps two stores per user that persist across sessions. The profile store holds personal facts and preferences, each with a validity period and a link to any record it replaces. The task store holds past tasks with their versions, results and files. Nothing enters the profile unless decision brain agrees, in a Choice question with a threshold of 0.8, that the speaker is explicitly stating a lasting fact about themselves or asking the assistant to remember one. Conditions that belong to a task stay on the task. In Figure~\ref{fig:teaser}, ``only sixty dollars tonight'' refines the snack task while ``keep forty as my usual limit'' is saved, and a new session opened later still answers that the usual budget is forty dollars. Only an explicit correction can replace a saved record, and the assistant says that something has been saved only after the write has been confirmed. The profile is loaded into every decision, and task records and older facts are retrieved on demand, with decision brain keeping at most three of twelve candidates (Section~\ref{sec:exp-memory}).

\section{Jarvis-Omni: When Evidence Expires}
\label{sec:omni}

Giving the assistant eyes changes the coordination problem in one important way. In audio, the facts behind a decision are mostly in the user's words, and words do not change after they are spoken. On a screen, facts change on their own. A player asks whether to push forward while a teammate is alive, and two seconds later the teammate is dead. A health bar that was low is full again after a recall. Advice that was right when the question was asked can be wrong by the time it is spoken. \sysomni is our audio-visual instance of \sysname, built on Venus-Omni~\cite{venus2026}. It treats visual facts the way \sysaudio treats tasks, as state with a version and a lifetime, and checks them again before anything based on them reaches the user.

The harness carries over almost unchanged. \sysomni uses the same \ledger, playback control and typed controller as \sysaudio, so tasks, versions, epochs and receipts behave identically, and the four principles of Section~\ref{sec:principles} apply as they are. Three parts are new, namely a perception layer that records what the system has seen and how long it stays true, a multimodal decision brain that can look at frames, and a gate that also decides when the screen itself deserves a comment.

\subsection{Evidence Timeline and Belief State}
\label{sec:belief}

All media enter an \timeline that keeps screen frames, game audio and the microphone on separate tracks aligned to one clock. Every decision reads a window that ends at the decision time, so no model ever sees a frame from the future. A new match or scene opens a new scene epoch and clears the window, so media from an earlier scene cannot leak into the current one.

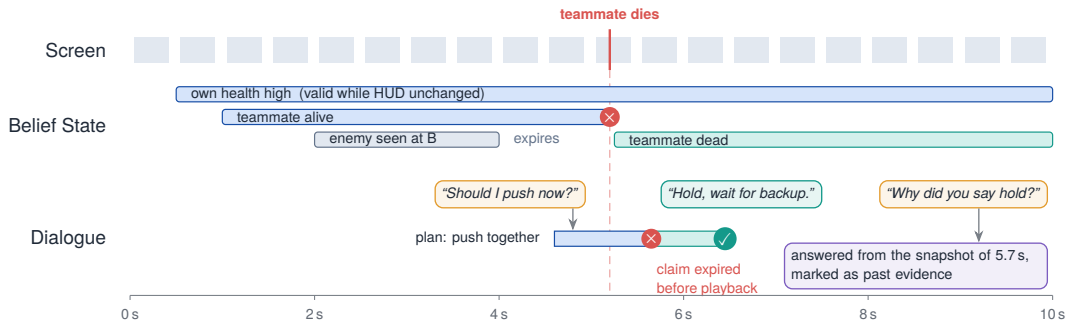
\begin{figure}[t]
    \centering
\begin{tikzpicture}[x=1.22cm, y=1cm, font=\sffamily\scriptsize]
  \tikzset{
    claim/.style={draw=jBlue, fill=jBlueSoft, rounded corners=1.2pt, line width=0.45pt},
    oldclaim/.style={draw=jSlate, fill=jSlateSoft, rounded corners=1.2pt, line width=0.45pt},
    newclaim/.style={draw=jTeal, fill=jTealSoft, rounded corners=1.2pt, line width=0.45pt},
    cl/.style={font=\sffamily\tiny, text=jInk},
    bub/.style={rounded corners=3pt, draw=#1, fill=#1!10, line width=0.5pt, font=\sffamily\tiny,
                text=jInk, align=left, inner sep=2.5pt},
    rowlab/.style={anchor=east, text=jInk, font=\sffamily\scriptsize},
    ar/.style={-{Stealth[length=1.6mm]}, line width=0.5pt, draw=jInk!60},
  }

  \node[rowlab] at (-0.15, 3.05) {Screen};
  \node[rowlab] at (-0.15, 2.05) {\belief};
  \node[rowlab] at (-0.15, 0.55) {Dialogue};

  \foreach \t in {0.0,0.5,...,9.5} {
    \fill[jSlateSoft] (\t+0.05, 2.88) rectangle (\t+0.42, 3.22);
  }
  \draw[jCoral, line width=0.9pt] (5.2, 2.78) -- (5.2, 3.32);
  \node[font=\sffamily\tiny\bfseries, text=jCoral, anchor=south] at (5.2, 3.32) {teammate dies};
  \draw[jCoral!60, densely dashed, line width=0.4pt] (5.2, 2.78) -- (5.2, -0.15);

  \draw[claim] (0.5, 2.36) rectangle (10, 2.56);
  \node[cl, anchor=west] at (0.55, 2.46) {own health high \ (valid while HUD unchanged)};
  \draw[claim] (1.0, 2.06) rectangle (5.2, 2.26);
  \node[cl, anchor=west] at (1.05, 2.16) {teammate alive};
  \node[circle, fill=jCoral, text=white, inner sep=0.6pt, font=\sffamily\tiny\bfseries] at (5.2, 2.16) {$\times$};
  \draw[oldclaim] (2.0, 1.76) rectangle (4.0, 1.96);
  \node[cl, anchor=west] at (2.05, 1.86) {enemy seen at B};
  \node[font=\sffamily\tiny, text=jSlate, anchor=west] at (4.05, 1.86) {expires};
  \draw[newclaim] (5.25, 1.76) rectangle (10, 1.96);
  \node[cl, anchor=west] at (5.3, 1.86) {teammate dead};

  \node[bub=jAmber, anchor=south west] (q) at (3.3, 0.95) {\textit{``Should I push now?''}};
  \draw[draw=jBlue, fill=jBlueSoft, line width=0.45pt] (4.6, 0.45) rectangle (5.65, 0.65);
  \node[cl, anchor=east] at (4.55, 0.55) {plan: push together};
  \draw[draw=jTeal, fill=jTealSoft, line width=0.45pt] (5.7, 0.45) rectangle (6.45, 0.65);
  \draw[ar] (4.8, 0.95) -- (4.8, 0.66);
  \node[circle, fill=jCoral, text=white, inner sep=0.6pt, font=\sffamily\tiny\bfseries] at (5.65, 0.55) {$\times$};
  \node[circle, fill=jTeal, text=white, inner sep=0.6pt, font=\sffamily\tiny\bfseries] at (6.45, 0.55) {\checkmark};
  \node[font=\sffamily\tiny, text=jCoral, anchor=north west, align=left] at (5.6, 0.36) {claim expired\\before playback};
  \node[bub=jTeal, anchor=south west] (a) at (5.75, 0.95) {\textit{``Hold, wait for backup.''}};

  \node[bub=jAmber, anchor=south east] (q2) at (9.95, 0.95) {\textit{``Why did you say hold?''}};
  \node[bub=jPlum, anchor=north east, text width=3.3cm] (past) at (9.95, 0.5) {answered from the snapshot of 5.7\,s,\\marked as past evidence};
  \draw[ar] (9.2, 0.95) -- (9.2, 0 |- past.north);

  \draw[jInk!60, line width=0.4pt] (0, -0.2) -- (10, -0.2);
  \foreach \t in {0,2,4,6,8,10} {
    \draw[jInk!60] (\t, -0.2) -- (\t, -0.26);
    \node[font=\sffamily\tiny, text=jInk!70, below] at (\t, -0.24) {\t\,s};
  }
\end{tikzpicture}
    \caption{Evidence that expires. Claims read from the screen carry a validity window and are invalidated by events such as a death or a camera change. Advice is checked against the \belief when it is ready, so a recommendation that rests on an expired claim is dropped and planned again from the new state. A question about earlier advice is answered from a historical snapshot.}
    \label{fig:belief}
    \end{figure}

What the system concludes from these media lives in a \belief of evidence claims, such as ``the teammate on the left is alive'' or ``my health is low''. Each claim records the observations it rests on, the time until which it is valid and the events that invalidate it, namely a change of scene, a change of camera, a lost source or a correction from the user. A claim may support a current decision only while it is valid and none of these events has occurred. When a response is ready, its claims are checked again. If one has expired, the response is discarded before playback, and if one expires while the response is playing, playback stops at that moment. Figure~\ref{fig:belief} shows how this unfolds over a few seconds of play.

Questions about the past, such as ``why did you tell me to retreat?'', are answered from a historical snapshot of the readings behind the advice, framed as what was on screen then, and a claim from a snapshot can never serve as current evidence.

\subsection{A Multimodal Decision Brain}
\label{sec:jevomni}

The decision brain of \sysomni is \jevomni, the multimodal member of the \jev family, a 12B typed judgment model built on Gemma~4~\cite{gemma4}. It answers the same Choice, Noul and Score questions as \jev, and its inputs may include causal still frames or a short causal audio window. Because the interface is the same, every decision of \sysaudio works in \sysomni without changes to the runtime. Two new decisions use the images. One reviews whether a frame supports a visual claim and whether that claim is still fresh, and the other chooses the tactical advice that fits the current screen. Game HUDs are dense and small, so \jevomni carries light adapters of about 8.5M parameters each for reading the HUD and for telling whether the player's character is alive. Open-ended descriptions of the screen are proposed by a separate vision-language model, Gemma-4-12B-IT~\cite{gemma4}\footnote{\url{https://huggingface.co/google/gemma-4-12B-it}}, but they must pass this review before they can support anything the assistant says.

Not every question needs the full slow brain. ``Should I push now?'' in the middle of a fight needs an answer within about two seconds. For such questions \jevomni reads the latest screenshot and picks the advice that fits it, an independent review confirms that the screen supports the advice, and the front model speaks it. In recorded sessions this fast path starts speaking about 2.2~s after the user finishes the question. Open questions, requests to remember something and questions about earlier rounds go through the full path with workers, tools and in-match memory (Appendix~\ref{app:omni}).

\subsection{A Gate That Also Watches}
\label{sec:omni-gate}

The gate of \sysomni follows the recipe of Section~\ref{sec:gate} with a richer causal input. It combines microphone and screen features with the dialogue so far, including the assistant audio that was actually played, and leaves out the transcript of the current utterance. It can also emit a fourth action, \act{SCENE}, which starts an analysis of the screen without any user request so the assistant can speak up when something important happens. We trained the rank-8 adapter with GRPO on 1,211 episodes built from 49 recorded game sessions. On a fixed test set of 96 episodes, the frozen control head of Venus-Omni makes the right decision in 30 cases and misses an action it should have taken in 66. Our adapter reaches 75 correct with a single missed action, against 67 and 22 for an earlier adapter without dialogue context. Missing almost nothing is what matters most here, because a request that the gate misses never reaches any later part of the system. In live sessions on the same user recordings, the new adapter also brings the first spoken response forward from 5.05~s to 1.78~s at the median.

\section{Experiments}
\label{sec:experiments}

Our experiments answer three questions. Does the harness by itself make a voice agent better, with everything else held fixed (\S\ref{sec:exp-paired})? How does the result compare with the strongest published voice systems (\S\ref{sec:exp-published})? And why does it work, mechanism by mechanism (\S\ref{sec:exp-mechanisms} and \S\ref{sec:exp-memory})? We close by asking whether the same ideas carry over to vision (\S\ref{sec:exp-omni}).

\subsection{Setup}
\label{sec:setup}

\paragraph{Benchmarks.}
$\tau$-Voice~\cite{tauvoice2026} runs the customer-service tasks of $\tau^2$-bench~\cite{tau2bench2025} as full-duplex voice calls with a simulated caller. We use all 278 regular tasks in airline, retail and telecom, and a task counts as solved only when the official reward, which checks the final database state and the required actions, equals 1. Full-Duplex-Bench v3~\cite{fullduplexbench_v3} (FDB-v3) plays 100 real recordings of people asking for multi-step tool use with natural disfluencies. Full-Duplex-Bench v2~\cite{fullduplexbench_v2} (FDB-v2) runs 200 multi-turn tasks with an automated examiner at two paces, and we report its task score from 1 to 5 for Correction, Entity tracking and Safety. ContextDialog~\cite{contextdialog2025} tests whether an assistant recalls earlier parts of a spoken conversation.

\paragraph{Front models and baselines.}
Besides Venus-Audio, we run the harness unchanged on MiniCPM-o~4.5~\cite{cui2026minicpm} and VoiceChat-11B~\cite{voicechat2026}. Our main baseline is Qwen Audio Agent (QAA)~\cite{qwen_audio_agent_2026}, an open voice runtime with a task manager, cancellation and playback receipts, which we run from its pinned upstream code. We also run Gander~\cite{gander2026}, MoshiRAG~\cite{moshirag2026} and the official harness released with Venus-Audio. Every harness uses the same DeepSeek-V4-Flash back end and budget, and all rows within a comparison share one judge (Appendix~\ref{app:protocol}).

\subsection{Does the Harness Matter?}
\label{sec:exp-paired}

Comparisons between published systems mix differences in front models, back ends and protocols. Our first experiment removes them. Table~\ref{tab:paired} keeps the front model, the admission path and the back end fixed and changes only the harness, so every difference comes from what happens to a turn after it is admitted.

\begin{table}[t]
\centering
\caption{Harnesses compared on the same front model, admission path and back end. $\tau$-Voice reports solved tasks, FDB-v3 reports passes out of 100 under exact and semantic matching, and FDB-v2 reports the mean task score over 300 task and pace pairs. All rows share one judge. $^\dagger$Released runtime with a DeepSeek-V4-Flash worker in place of its Codex back end.}
\label{tab:paired}
\tablefont
\fullwidthtabular{llccccccc}{%
\toprule
\tableheader & & \multicolumn{4}{c}{\textbf{$\tau$-Voice} (solved)} & \multicolumn{2}{c}{\textbf{FDB-v3}} & \textbf{FDB-v2} \\
\tableheader \textbf{Front model} & \textbf{Harness} & Airline & Retail & Telecom & Total & Exact & Semantic & Task score \\
\cmidrule(lr){3-6} \cmidrule(lr){7-8} \cmidrule(lr){9-9}
\multirow{3}{*}{Venus-Audio} & Native harness & 17 & 6 & 0 & 23 & 45 & 35 & 3.09 \\
 & QAA & 23 & 6 & 0 & 29 & 41 & 59 & 2.57 \\
 & \cellcolor{oursRow}\textbf{\sysname} & \cellcolor{oursRow}\textbf{35} & \cellcolor{oursRow}\textbf{77} & \cellcolor{oursRow}\textbf{80} & \cellcolor{oursRow}\textbf{192} & \cellcolor{oursRow}\textbf{51} & \cellcolor{oursRow}\textbf{63} & \cellcolor{oursRow}\textbf{3.36} \\
\midrule
\multirow{2}{*}{MiniCPM-o 4.5} & QAA & 23 & 6 & 0 & 29 & 40 & 56 & 2.98 \\
 & \cellcolor{oursRow}\textbf{\sysname} & \cellcolor{oursRow}\textbf{23} & \cellcolor{oursRow}\textbf{6} & \cellcolor{oursRow}\textbf{9} & \cellcolor{oursRow}\textbf{38} & \cellcolor{oursRow}\textbf{50} & \cellcolor{oursRow}\textbf{63} & \cellcolor{oursRow}\textbf{3.28} \\
\midrule
\multirow{2}{*}{VoiceChat-11B} & QAA & 1 & 1 & 0 & 2 & 26 & 38 & 2.44 \\
 & \cellcolor{oursRow}\textbf{\sysname} & \cellcolor{oursRow}\textbf{23} & \cellcolor{oursRow}\textbf{6} & \cellcolor{oursRow}\textbf{4} & \cellcolor{oursRow}\textbf{33} & \cellcolor{oursRow}\textbf{34} & \cellcolor{oursRow}\textbf{50} & \cellcolor{oursRow}\textbf{2.84} \\
\midrule
\tablegroup{9}{Other full-duplex systems with their own runtimes} \\
MoshiRAG & Native retrieval & 0 & 0 & 0 & 0 & 0 & 0 & 3.08 \\
Gander & Native runtime$^\dagger$ & 23 & 6 & 0 & 29 & 5 & 6 & 3.16 \\
\bottomrule}
\end{table}

\begin{figure}[t]
\centering
\includegraphics[width=\textwidth]{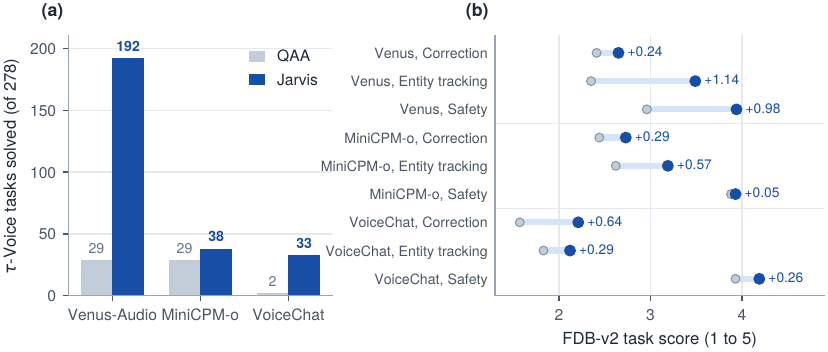}
\caption{\sysname against QAA with the same front model, admission path and back end. (a) $\tau$-Voice tasks solved out of 278. (b) FDB-v2 task score for every combination of front model and category. \sysname improves on QAA in all nine.}
\label{fig:paired}
\end{figure}

The harness turns out to matter enormously. On Venus-Audio, \sysname solves 192 of 278 $\tau$-Voice tasks against 29 for QAA. It passes 51 FDB-v3 recordings under exact matching against 41, a gain of 10 points with a paired bootstrap interval of $[3.0, 17.8]$, and raises the FDB-v2 task score from 2.57 to 3.36, with an interval of $[0.53, 1.04]$ for the gain. The same picture holds on the other two front models (Figure~\ref{fig:paired}). \sysname never trails QAA on any benchmark, domain or task category, and on VoiceChat it lifts $\tau$-Voice from 2 to 33 solved tasks. Systems that keep their own runtimes fall far behind on the two tool-use benchmarks under the same protocol. MoshiRAG retrieves knowledge but cannot execute tools, so it solves no $\tau$-Voice task, the official Venus harness passes no FDB-v3 recording, and Gander passes 5. On FDB-v2, which needs no tools, their task scores lie between 3.08 and 3.16, still below the 3.36 of \sysname on Venus-Audio.
The gains come from three places. First, fewer requests are lost. With identical admissions on Venus-Audio, QAA produces an audible response in 68 of the 100 FDB-v3 recordings and \sysname in 97, because the \ledger and \delivery make sure that an admitted request is answered, asked about or reported as failed, and never silently dropped. Second, arguments and answers are more accurate. Argument accuracy and response quality improve on all three front models, since the worker plans with the full policy and decision brain keeps a short reply such as a name or a number attached to the service it belongs to. Third, the gains concentrate on work that changes state. The hardest $\tau$-Voice tasks must write to the database, such as rebooking a flight or changing an order. QAA solves none of them on any front model, while \sysname solves 137 on Venus-Audio. None of this costs time. The median task-completion time on FDB-v3 is 12.9~s for \sysname against 14.2~s for QAA. Table~\ref{tab:fdb3-paired} in the appendix gives the full breakdown.

\begin{takeaway}
With the front model, admission and back end held fixed, the harness alone moves $\tau$-Voice success on Venus-Audio from 29 to 192 of 278 tasks. \sysname improves on QAA for all three front models and never falls behind it on any benchmark, domain or category.
\end{takeaway}

\subsection{Comparison with Leading Voice Systems}
\label{sec:exp-published}

\begin{table}[t]
\centering
\caption{Results on Full-Duplex-Bench v3 (100 recordings). The first four metrics are fractions and the last three are percentages. Bold marks the best value in each column, and -- means not reported.}
\label{tab:fdb3-published}
\tablefont
\setlength{\tabcolsep}{5pt}
\adjustbox{max width=\textwidth}{%
\begin{tabular}{lccccccc}
\toprule
\tableheader \textbf{System} & \textbf{Tool F1}$\uparrow$ & \textbf{Arg.\ acc.}$\uparrow$ & \textbf{Resp.\ qual.}$\uparrow$ & \textbf{Pass@1}$\uparrow$ & \textbf{Take-turn}$\uparrow$ & \textbf{Interrupt}$\downarrow$ & \textbf{Filler}$\downarrow$ \\
\midrule
\tablegroup{8}{Commercial and cascaded systems} \\
GPT-Realtime & 0.876 & 0.680 & 0.792 & 0.600 & 96.0 & 13.5 & 16.9 \\
Gemini Live 3.1 & 0.817 & 0.588 & 0.718 & 0.540 & 78.0 & 19.2 & 31.7 \\
Gemini Live 2.5 & 0.786 & 0.593 & 0.554 & 0.490 & 92.0 & 14.1 & \textbf{8.9} \\
Cascaded (Whisper, GPT-4o, TTS) & 0.803 & 0.562 & 0.600 & 0.450 & \textbf{100.0} & 33.0 & 26.9 \\
Grok & 0.797 & 0.542 & 0.617 & 0.430 & 94.0 & 25.5 & 44.3 \\
Ultravox v0.7 & 0.794 & 0.513 & 0.510 & 0.410 & 96.0 & 47.9 & 88.0 \\
\midrule
\tablegroup{8}{Open full-duplex models and agents} \\
Venus-Audio & 0.850 & 0.542 & -- & 0.480 & -- & -- & -- \\
Venus-Omni & 0.860 & 0.531 & -- & 0.430 & -- & -- & -- \\
VoiceChat-11B & 0.825 & 0.422 & -- & 0.330 & -- & -- & -- \\
Gander & 0.759 & 0.503 & 0.490 & 0.400 & \textbf{100.0} & 8.0 & 51.6 \\
\midrule
\oursrow \textbf{\sysaudio} & \textbf{0.927} & \textbf{0.783} & \textbf{0.850} & \textbf{0.690} & \textbf{100.0} & \textbf{7.0} & 20.4 \\
\bottomrule
\end{tabular}%
}
\end{table}

Table~\ref{tab:fdb3-published} compares \sysaudio with the systems reported on FDB-v3. \sysaudio reaches a Pass@1 of 69\%, nine points above GPT-Realtime and 21 points above the Venus-Audio model it runs on. It also leads on tool selection, argument accuracy and response quality, takes the turn in every recording and interrupts the user in only 7\% of them, the lowest rate in the table. The margin grows with difficulty. On hard scenarios \sysaudio passes 56.7\% against 43.3\% for GPT-Realtime (Table~\ref{tab:fdb3-breakdown} in the appendix).

\begin{table}[t]
\centering
\caption{Task success (\%) on $\tau$-Voice (regular tasks). Bold marks the best value among systems without a proprietary reasoning back end.}
\label{tab:tau-published}
\tablefont
\fullwidthtabular{lcccc}{%
\toprule
\tableheader \textbf{System} & \textbf{Retail}$\uparrow$ & \textbf{Airline}$\uparrow$ & \textbf{Telecom}$\uparrow$ & \textbf{Macro avg.}$\uparrow$ \\
\midrule
\tablegroup{5}{Realtime speech models} \\
GPT-Realtime 1.0 & 35.96 & 36.00 & 19.30 & 30.42 \\
GPT-Realtime 1.5 & 44.70 & 40.00 & 21.10 & 35.27 \\
GPT-Realtime 2 & 47.37 & 58.00 & 21.93 & 42.43 \\
Gemini 2.5 Flash Live & 29.80 & 30.00 & 17.50 & 25.77 \\
Gemini 3.1 Flash Live (thinking high) & 45.61 & 64.00 & 21.93 & 43.85 \\
Grok Voice Fast 1.0 & 38.60 & 36.00 & 40.40 & 38.33 \\
Grok Voice Think Fast 1.0 & 62.28 & 66.00 & \textbf{73.68} & 67.32 \\
Grok Voice Think Fast 2.0 & 59.65 & 56.00 & 71.93 & 62.53 \\
Qwen3.5-Omni-Plus Realtime & 46.49 & 54.00 & 60.53 & 53.67 \\
\midrule
\tablegroup{5}{Cascaded pipeline} \\
LiveKit (Nova-3, GPT-4.1, Aura) & 28.95 & 48.00 & 16.67 & 31.20 \\
\midrule
\tablegroup{5}{Realtime front end with a proprietary reasoning back end} \\
GPT Live 1 with GPT-6 Astra & 78.95 & 82.00 & 84.21 & 81.72 \\
Pine Voice Preview & 70.18 & 70.00 & 85.96 & 75.38 \\
\midrule
\oursrow \textbf{\sysaudio} (open 9B front, DeepSeek-V4-Flash) & \textbf{67.54} & \textbf{70.00} & 70.18 & \textbf{69.24} \\
\bottomrule}
\end{table}

On the public $\tau$-Voice leaderboard~\cite{tauleaderboard2026} (Table~\ref{tab:tau-published}), \sysaudio reaches a macro success rate of 69.2\%, higher than every realtime speech model on the board, including GPT-Realtime~2 at 42.4\% and Grok Voice Think Fast 1.0 at 67.3\%, and more than double the cascaded pipeline. The two systems ranked above it, GPT Live~1 and Pine Voice Preview, follow the same fast and slow design but rely on proprietary frontier models in the background. \sysaudio gets close with an open 9B front model and a flash-tier back end, which again suggests that how the two brains are coordinated matters as much as how strong the slow brain is.

\subsection{Why It Works}
\label{sec:exp-mechanisms}

A benchmark caller rarely changes a request while the work on it is still running, so we test the mechanisms directly.

\paragraph{Each ledger check removes a different failure.}
\label{sec:exp-race}
We built a race suite in which twenty templates with four timing variants each inject a refinement, a cancellation, a correction during planning, a switch to another task or an interruption during playback. Each of the 80 scripts runs three times per configuration against the frozen runtime of \sysaudio with its real tool loop and native speech, and the script fixes the control decisions so that the runtime is tested on its own. A run is consistent only if no stale audio, stale tool effect or false delivery record occurs and every result that should be delivered is played in full.

\begin{table}[t]
\centering
\caption{Consistency under 80 race scripts, each run three times per configuration. The middle columns count runs with each error. A false delivery is a task recorded as delivered although the user never heard all of it. The last column gives the paired gain of the full runtime in consistent runs, with a 95\% bootstrap interval clustered by template.}
\label{tab:stress}
\tablefont
\setlength{\tabcolsep}{5pt}
\adjustbox{max width=\textwidth}{%
\begin{tabular}{lccccc}
\toprule
\tableheader \textbf{Runtime} & \textbf{Consistent runs}$\uparrow$ & \textbf{Stale audio}$\downarrow$ & \textbf{Stale tool effect}$\downarrow$ & \textbf{False delivery}$\downarrow$ & \textbf{Gain of full runtime (pp)} \\
\midrule
\oursrow \textbf{\sysname, full runtime} & \textbf{240 / 240} & \textbf{0} & \textbf{0} & \textbf{0} & \textendash \\
\quad without result validity (I2) & 168 / 240 & 72 & 0 & 0 & $+30.0$ \ $[15.0, 45.0]$ \\
\quad without plan freshness (I3) & 168 / 240 & 72 & 72 & 0 & $+30.0$ \ $[15.0, 45.0]$ \\
\quad without playback receipts (I4) & 204 / 240 & 0 & 0 & 36 & $+15.0$ \ $[3.8, 30.0]$ \\
\bottomrule
\end{tabular}%
}
\end{table}

The full runtime is consistent in all 240 runs (Table~\ref{tab:stress}). Without the result check of invariant~I2, an answer from a superseded or cancelled version reaches the speaker in 72 runs. Without the plan check of invariant~I3, the outdated plan also executes its tools in the same 72 runs, so the user hears a wrong answer and the wrong action has already been taken. Without playback receipts, 36 runs end with a task marked as delivered that the user never heard in full. No check blocks a legitimate delivery, since all configurations pass the 60 control runs in which the work should continue.

\paragraph{JEV decides correctly, and in time.}
\label{sec:exp-jev}
The checks act on decisions, and a decision helps only if it arrives in time. A cancellation recognized after the old worker has finished is too late, because the result is already on its way to the speaker. We compare \jev with two general LLMs used as controllers, DeepSeek-V4-Flash and Gemini-3.5-Flash. All three receive the same options and context, and for each we pick on 240 development cases the threshold that accepts the most control operations while keeping wrong controls at no more than 1\%. The test set holds 192 new bilingual cases from 24 scenario families, including cases where a domain action must be told apart from a harness control, and every controller sees each case three times.

\begin{table}[t]
\centering
\caption{\jev against general LLM controllers with the same options and context. (a)~Decisions on 192 test cases, each called three times. In time counts correct decisions that clear the threshold within 1~s, and with fallback also counts the conservative \op{ANSWER} fallback within the 3.5~s deadline of the runtime. (b)~The measured decisions replayed in the runtime on 64 cases at four worker completion times. A correct binding applies the right operation to the right task and changes no other task, and stopped in time counts cancellations and refinements applied before the old worker finished.}
\label{tab:abl-jev}
\tablefont
\textbf{(a) Decisions on the test set} \\[3pt]
\fullwidthtabular{lcccc}{%
\toprule
\tableheader \textbf{Controller} & \textbf{Threshold} & \textbf{Median request (s)}$\downarrow$ & \textbf{In time, 1~s}$\uparrow$ & \textbf{With fallback, 3.5~s}$\uparrow$ \\
\midrule
DeepSeek-V4-Flash & 0.95 & 1.57 & 0 / 576 \phantom{(00.0\%)} & 465 / 576 (80.7\%) \\
Gemini-3.5-Flash & 1.00 & 3.14 & 0 / 576 \phantom{(00.0\%)} & 456 / 576 (79.2\%) \\
\oursrow \textbf{\jev} & 0.43 & \textbf{0.30} & \textbf{524 / 576 (91.0\%)} & \textbf{544 / 576 (94.4\%)} \\
\bottomrule}

\vspace{8pt}
\textbf{(b) Measured decisions replayed in the runtime} \\[3pt]
\fullwidthtabular{lcccc}{%
\toprule
\tableheader \textbf{Controller} & \textbf{Correct binding}$\uparrow$ & \textbf{Stopped in time}$\uparrow$ & \textbf{Stale audio}$\downarrow$ & \textbf{Unchanged tasks kept}$\uparrow$ \\
\midrule
DeepSeek-V4-Flash & 220 / 256 & 77 / 224 & 36 & 32 / 32 \\
Gemini-3.5-Flash & 160 / 256 & \phantom{0}9 / 224 & 96 & 32 / 32 \\
\oursrow \textbf{\jev} & \textbf{252 / 256} & \textbf{150 / 224} & \textbf{4} & 32 / 32 \\
\bottomrule}
\end{table}

\begin{figure}[t]
\centering
\includegraphics[width=\textwidth]{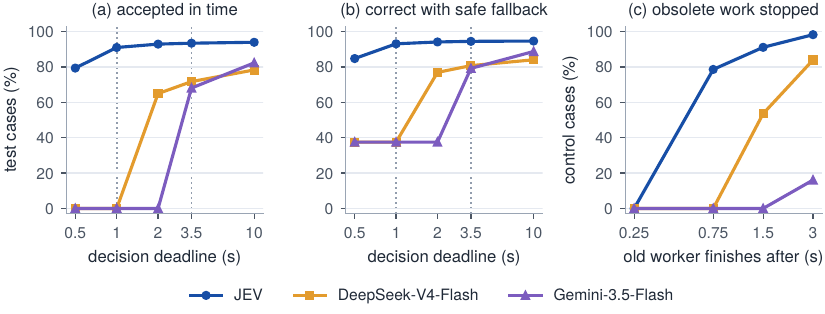}
\caption{Timely control. (a) Correct decisions that clear each controller's threshold within a given deadline. (b) Correct decisions when the conservative fallback is counted. (c) Share of cancellations and refinements that stop the old worker before it finishes, as a function of when it would finish. Dotted lines mark the 1~s window and the 3.5~s deadline of the runtime.}
\label{fig:jev-deadlines}
\end{figure}

\jev returns a correct and confident decision within one second in 91\% of the 576 test calls (Table~\ref{tab:abl-jev}a). Neither LLM controller does so even once, because their median requests take 1.57~s and 3.14~s against 0.30~s for \jev. Even at the 3.5~s deadline with the conservative fallback counted, \jev is correct in 94.4\% of the calls against about 80\% for both, and none of the control operations it accepted was wrong. Speed matters because the runtime acts on it. When we replay the measured decisions in the real runtime with the old worker finishing between 0.25 and 3~s after the turn, \jev binds the right operation to the right task in 252 of 256 replays and stops the obsolete worker in time in 150 of 224 cases (Table~\ref{tab:abl-jev}b). The two LLM controllers stop it in 77 and 9 cases, and stale speech reaches the speaker in 36 and 96 of their replays against 4 for \jev. Figure~\ref{fig:jev-deadlines}(c) shows where the difference comes from. When the old worker finishes 0.75~s after the turn, \jev still stops 79\% of the obsolete work and neither LLM controller stops any.

\begin{takeaway}
Each ledger check removes a different failure, and together they keep the runtime consistent in all 240 race runs. \jev supplies the decisions these checks act on, and it supplies them in time. It is correct and confident within one second in 91\% of the test cases, which no general LLM controller achieves, and in the runtime it stops about twice as much obsolete work as the better of them.
\end{takeaway}

\subsection{Memory}
\label{sec:exp-memory}

\begin{table}[t]
\centering
\caption{Spoken conversational memory on ContextDialog (80 fixed questions, Venus-Audio). Scores are on a 1 to 5 scale after a uniform ASR audit of the played answers. Evidence is the mean number of tokens passed to the answering model.}
\label{tab:memory}
\tablefont
\setlength{\tabcolsep}{7pt}
\adjustbox{max width=\textwidth}{%
\begin{tabular}{lccc}
\toprule
\tableheader \textbf{Harness and memory} & \textbf{Mean score}$\uparrow$ & \textbf{Score $\ge 4$}$\uparrow$ & \textbf{Evidence tokens} \\
\midrule
QAA, default Markdown memory with automatic injection & 1.45 & 6 / 80 & 55 \\
\oursrow \textbf{\sysname}, \jev selection & 4.19 & 61 / 80 & 700 \\
\oursrow \textbf{\sysname}, hybrid dense and sparse retrieval & \textbf{4.41} & \textbf{65 / 80} & 1013 \\
\bottomrule
\end{tabular}%
}
\end{table}

We ask 80 fixed questions about 20 complete ContextDialog conversations, feed the history as real audio and score the answer that was actually played. Both memory back ends of \sysname score close to 3 points higher than the default memory of QAA (Table~\ref{tab:memory}), with a conversation-clustered interval of $[2.43, 3.05]$ for the gain of \jev selection. The difference lies in what is stored. QAA compresses the conversation into short notes and passes only 55 tokens of evidence, which loses the details the questions ask about, while \sysname keeps what both speakers said. The two back ends differ in how much evidence they pass on. \jev selection keeps only the records it judges relevant, while the hybrid retriever passes on more of what it finds. On the 45 questions where \jev keeps some of the records, it scores 4.69 against 4.36 for the hybrid retriever, because a short and focused context is easier to answer from. Its overall mean is lower because on eight questions it discarded every record and left the answering model with no evidence.

\subsection{Does It Carry Over to Vision?}
\label{sec:exp-omni}

\begin{table}[t]
\centering
\caption{Visual input in the complete interaction path of \sysomni. A correct answer must be accepted by two blind automated reviews of the played audio and the recorded screen, and every question stays in the denominator. Silence counts controls in which the assistant correctly stays quiet. Intervals are 95\% bootstrap intervals clustered by upload.}
\label{tab:abl-omni}
\tablefont
\setlength{\tabcolsep}{7pt}
\adjustbox{max width=\textwidth}{%
\begin{tabular}{lccccc}
\toprule
\tableheader \textbf{Cohort} & \textbf{Uploads} & \textbf{With screen}$\uparrow$ & \textbf{Without screen} & \textbf{Gain (pp)} & \textbf{Silence kept}$\uparrow$ \\
\midrule
\oursrow First, full & 16 & \textbf{71 / 300 (23.7\%)} & 3 / 300 (1.0\%) & $+22.7$ \ $[18.8, 25.6]$ & 60 / 60 \\
\oursrow Second, compact & 12 new & \textbf{30 / 120 (25.0\%)} & 3 / 120 (2.5\%) & $+22.5$ \ $[14.2, 30.8]$ & 24 / 24 \\
\bottomrule
\end{tabular}%
}
\end{table}

\paragraph{The answers rest on the screen.}
We ran \sysomni end to end on recorded game sessions, with the real browser, speech recognition, visual processing and native speech output. Each recording comes with six conditions, such as two different screens for the same question, a change in the game state and a silent control, and the ablation removes only the screen frames that enter the runtime. An answer counts as correct only if two blind reviews of the played audio and the recorded screen both accept it, and questions answered late or not at all stay in the denominator. In two independent cohorts, removing the screen drops correct answers from 23.7\% to 1.0\% and from 25.0\% to 2.5\%, gains of 22.7 and 22.5 points (Table~\ref{tab:abl-omni}). Almost every correct answer therefore rests on visual evidence, and both configurations stay quiet in every silent control.

\paragraph{Expired evidence stays off the speaker.}
An answer that was right when it was admitted can still expire while it waits in the speech queue or while it plays. We played admitted utterances through the real speech scheduler and browser audio player, with their evidence set to expire either before or during playback. Without the validity check, 12.3~s of expired audio reaches the user when the evidence expires in the queue and 9.9~s when it expires during playback. With the check, these fall to zero and to 0.125~ms, three samples in total, while utterances whose evidence stays valid play in full. The rule that keeps a stale task result off the speaker in \sysaudio thus carries over to visual evidence, all the way to the sound the user hears.

\begin{takeaway}
The answers of \sysomni rest on the screen. Removing it costs about 22 points of correct answers in two independent cohorts, and silent matches stay silent either way. The validity rule follows the evidence through the speech queue into playback, so expired advice is not spoken.
\end{takeaway}

\section{Jarvis in Action}
\label{sec:applications}

\begin{figure}[t]
\centering
\captionsetup[sub]{justification=raggedright}
\begin{subfigure}[t]{0.4772\textwidth}
\centering
\includegraphics[width=\linewidth]{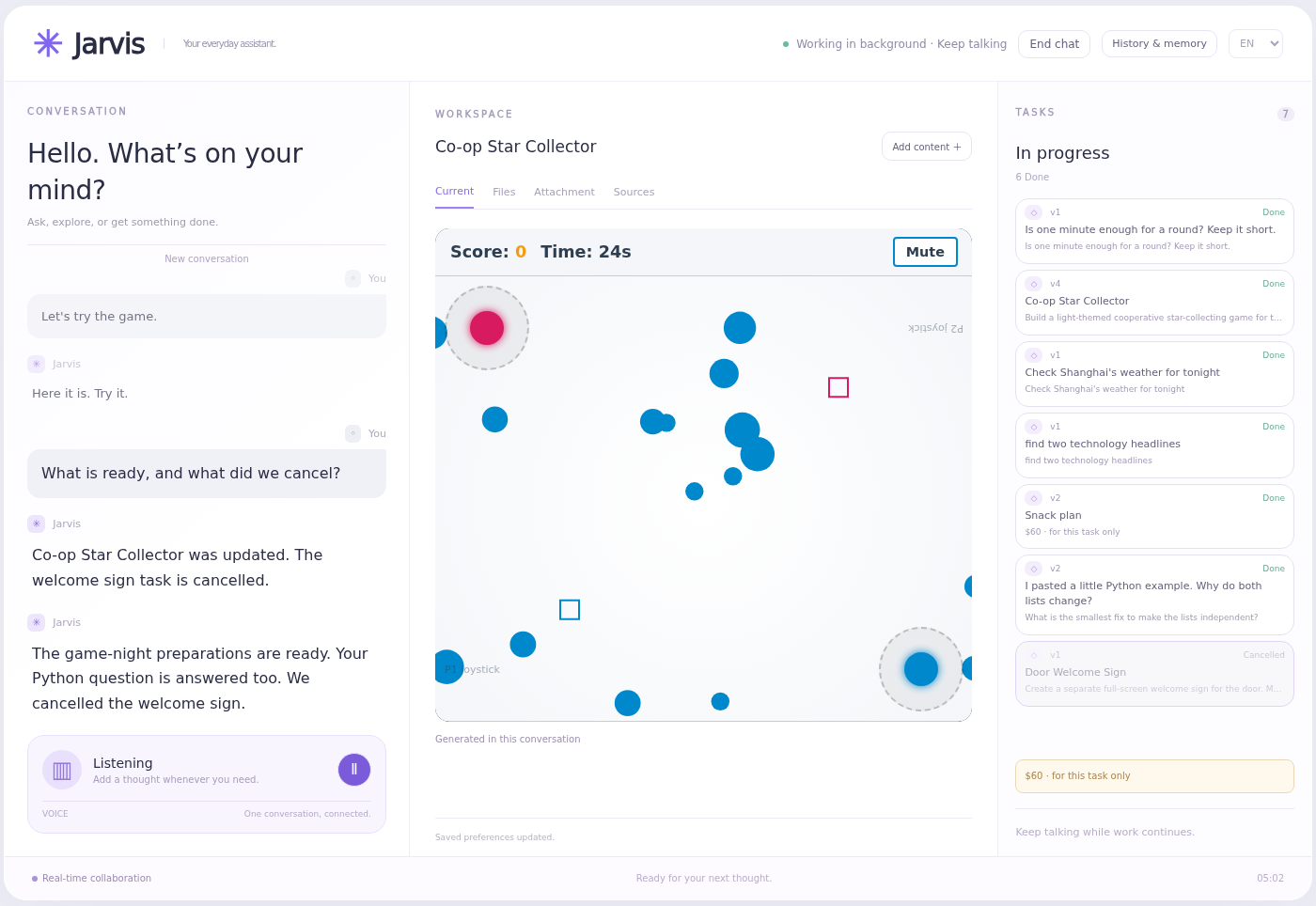}
\caption{The game-night session at its end. Each card carries its task version, and the welcome sign is the only cancelled task.}
\end{subfigure}\hfill
\begin{subfigure}[t]{0.4928\textwidth}
\centering
\includegraphics[width=\linewidth]{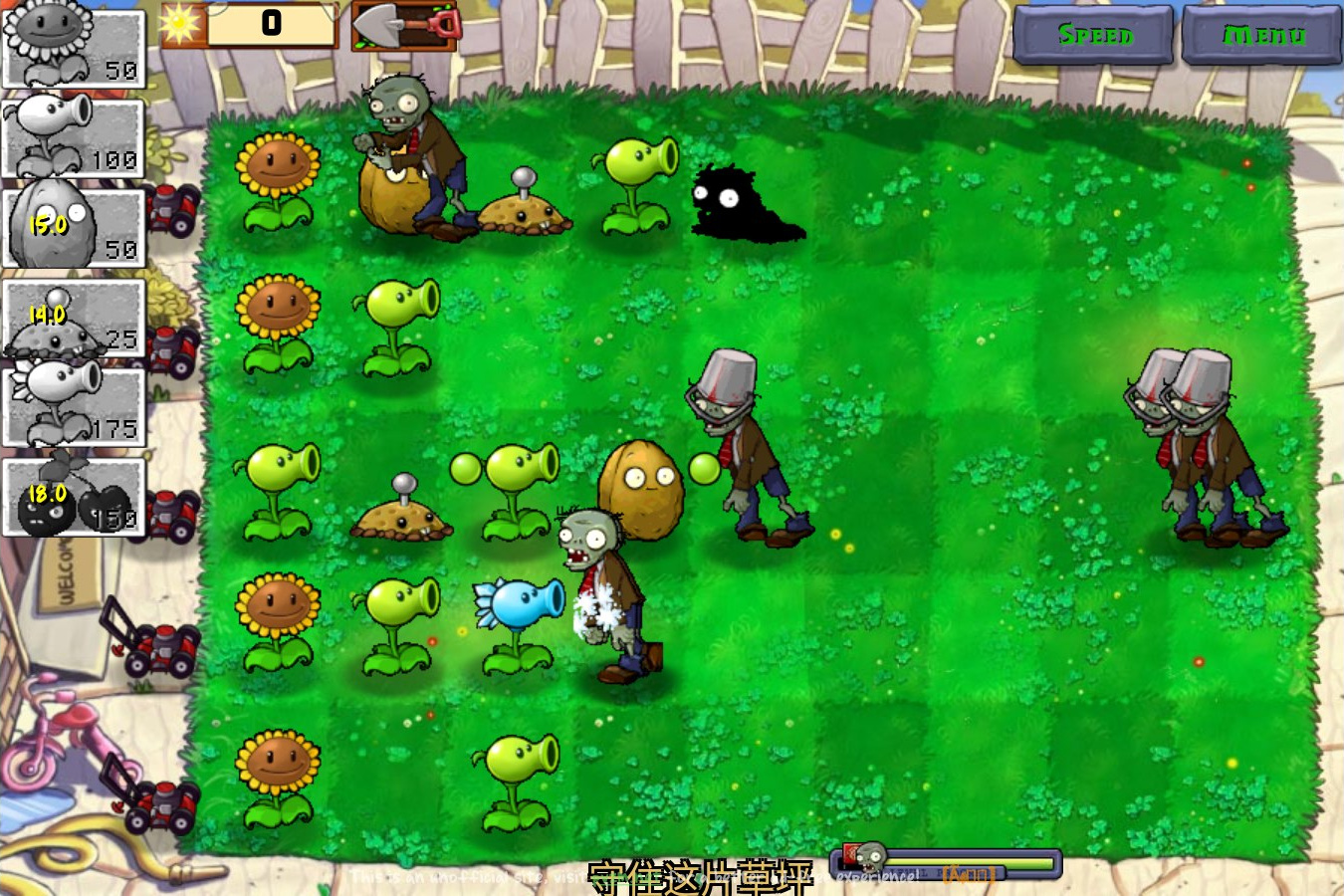}
\caption{Plants vs.\ Zombies. ``Lane 1 is under pressure. Use the cherry bomb here?'' The user can take over and hand back control at any time.}
\end{subfigure}

\vspace{6pt}
\begin{subfigure}[t]{0.2618\textwidth}
\centering
\includegraphics[width=\linewidth]{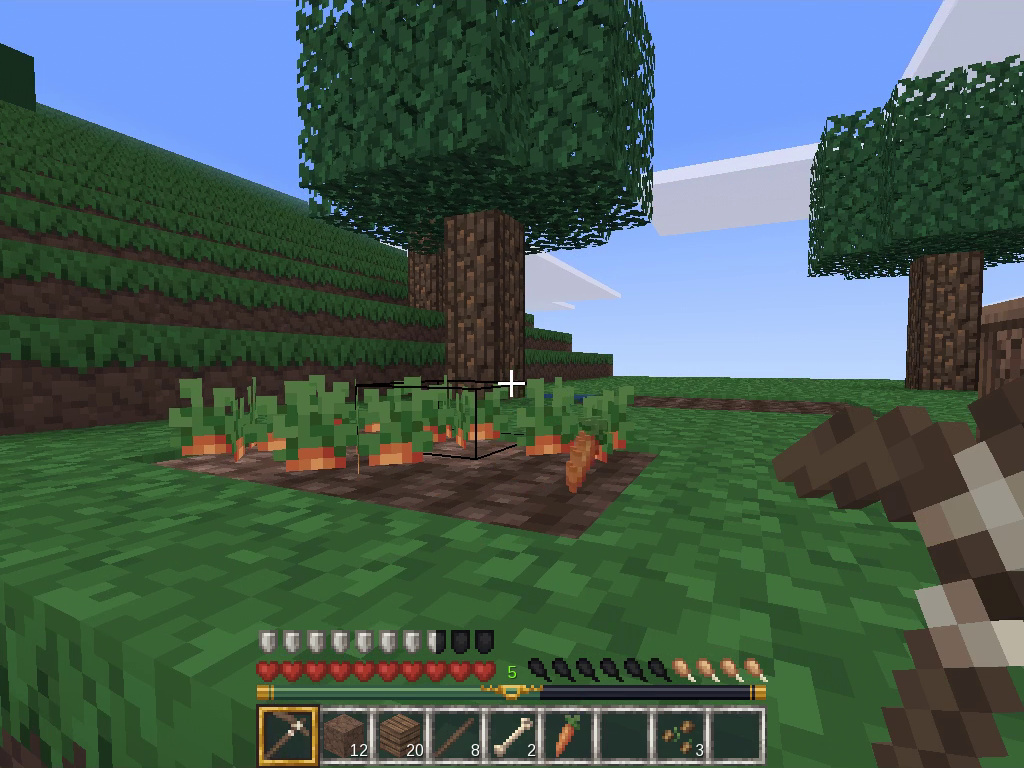}
\caption{VoxeLibre. Actions run under a revocable grant.}
\end{subfigure}\hfill
\begin{subfigure}[t]{0.3491\textwidth}
\centering
\includegraphics[width=\linewidth]{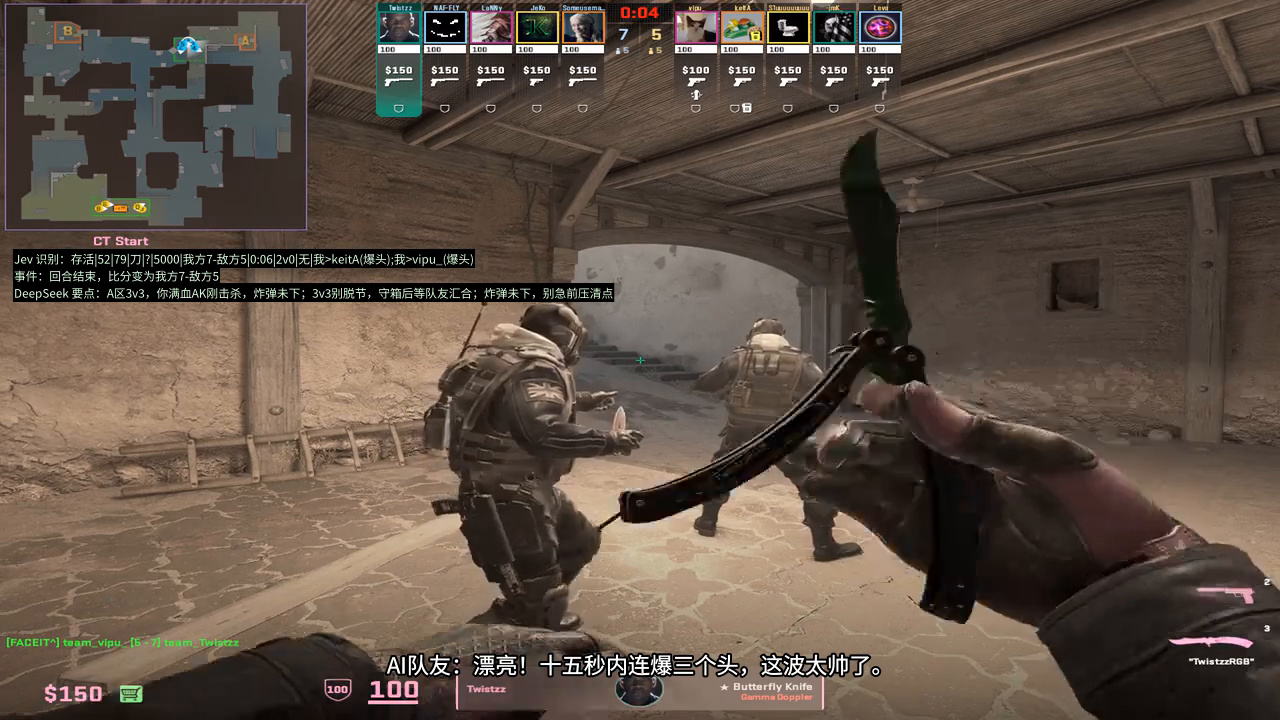}
\caption{CS2. ``Nice! Three headshots in fifteen seconds'', said on its own initiative.}
\end{subfigure}\hfill
\begin{subfigure}[t]{0.3491\textwidth}
\centering
\includegraphics[width=\linewidth]{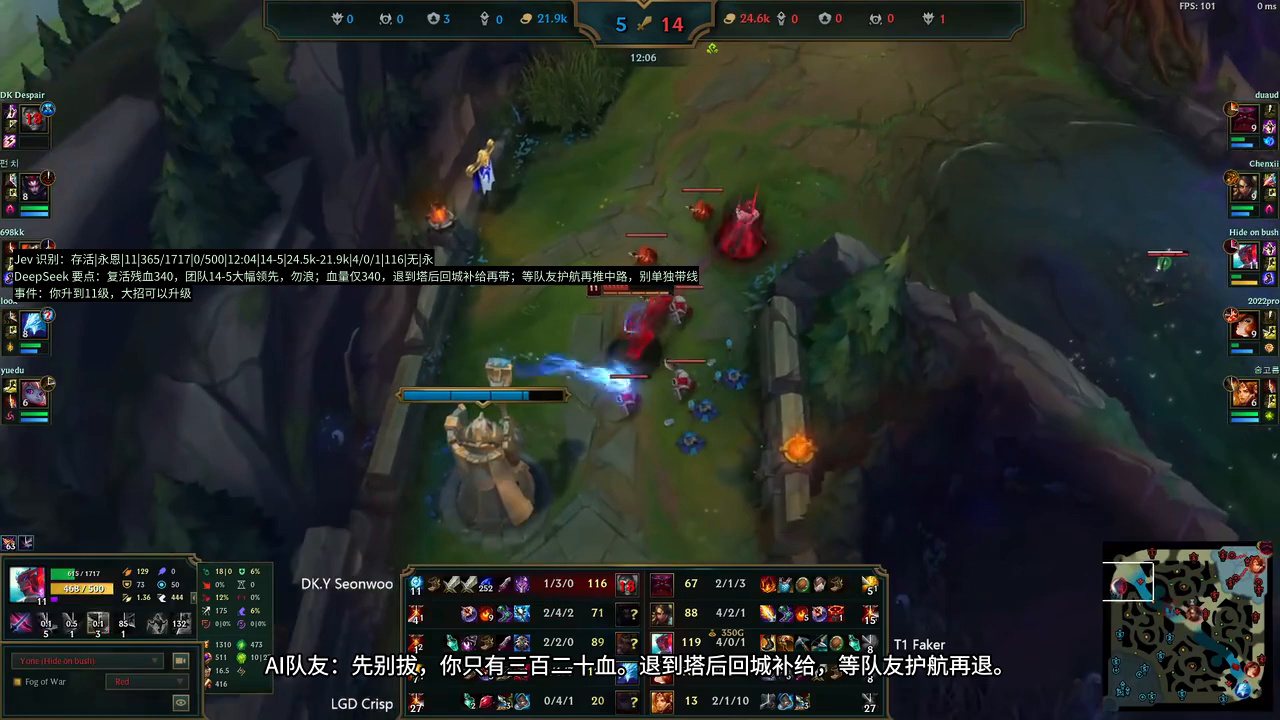}
\caption{League of Legends. Asked about taking the top tower: ``Hold off, you only have 320~HP''. Tactical advice in about 2.2~s.}
\end{subfigure}
\caption{\sysname in use. (a) to (c) run \sysaudio and (d) and (e) run \sysomni.}
\label{fig:cases}
\end{figure}

Benchmarks test parts of the problem one at a time. The sessions in Figure~\ref{fig:cases} show the whole system at work.

\paragraph{Daily assistance.}
At the end of the game night, the web client of Figure~\ref{fig:cases}(a) shows seven task cards with the version of each result, the cancelled welcome sign set apart and a note that the \$60 budget applies to tonight only. When the same user opens a new session later and asks about defaults, \sysname answers that forty dollars is the usual snack budget and that light themes are the saved preference for games.

\paragraph{Playing alongside the user.}
In Plants vs.\ Zombies, \sysaudio discusses strategy, keeps 150 sun in reserve because the user asked it to, and speaks up on its own when a lane comes under pressure (``Lane 1 is under pressure. Use the cherry bomb here?''). The user takes over and later hands control back, and the 5.3-minute session ends in a victory. In the voxel sandbox Luanti~\cite{luanti} with the VoxeLibre game~\cite{voxelibre}, it equips a sword, fights a husk and pauses its remaining actions when the user grabs the controls in the middle of the fight. Every move in both games goes through a revocable grant tied to the current game state, and every reported outcome, from each of the 23 plantings in Plants vs.\ Zombies to a harvested carrot, is a receipt from the game engine rather than a claim by the model.

\paragraph{Game companion.}
\sysomni watches recorded CS2 and League of Legends matches and talks about them. Asked what weapon a player was holding a moment ago, it answers from the earlier frame. Told to watch quietly and speak up at a highlight, it waits and then reacts to the victory screen with ``That round was a great win''. When the user asks for relaxed chat instead of constant tactics and asks it to remember that, it confirms the preference and switches to casual conversation.

\section{Related Work}
\label{sec:related}

\paragraph{Full-duplex spoken dialogue.}
Spoken dialogue has moved from turn-based pipelines to models that listen and speak at once, from dGSLM~\cite{nguyen2023dgslm} and SyncLLM~\cite{veluri2024syncllm} to Moshi~\cite{defossez2024moshi}, Freeze-Omni~\cite{freezeomni2024}, MiniCPM-o~4.5~\cite{cui2026minicpm}, Seeduplex~\cite{seeduplex2026}, BayLing-Duplex~\cite{BayLing-Duplex} and Realtime-Venus~\cite{venus2026}, with omni models extending the idea to video~\cite{qwen3omni,qwen3.5omni,seedrealtime2026}. Benchmarks have moved from turn-taking~\cite{fullduplexbench_v1} to instruction following~\cite{fullduplexbench_v2} and tool use~\cite{fullduplexbench_v3,tauvoice2026}. These models settle when to speak, and \sysname builds on them to decide what to do with an utterance once the system acts on it.

\paragraph{Fast and slow voice agents.}
Many recent systems pair a realtime model with a stronger background model~\cite{Mind-paced-speaking,huang2026duplexomni,thinkingmachines2026interactionmodels,openai_gpt_live_system_card_2026}. Realtime-Venus delegates through a native token~\cite{venus2026}, Qwen Audio Agent~\cite{qwen_audio_agent_2026} offers a runtime with a task manager and playback receipts, Gander~\cite{gander2026} lets the front model emit task calls to a Brain, and MoshiRAG~\cite{moshirag2026} retrieves knowledge while Moshi keeps talking. \sysname shares the split but differs in where coordination is decided. A dedicated decision brain selects among operations bound to task versions, and the hand-off is learned on the front model's native control branch instead of being prompted.

\paragraph{Dialogue state, typed decisions and memory.}
Task-oriented dialogue has long tracked explicit slot states~\cite{budzianowski2018multiwoz,wu2019trade}. \sysname tracks asynchronous tasks with versions and delivery status instead, updated while the work is still running. Tool-using models~\cite{yao2022react,schick2023toolformer,patil2024gorilla} and LLM judges~\cite{zheng2023judging} show that models can choose actions and grade outputs, and \jev turns this into distributions over runtime-supplied options that are easy to threshold and audit. Agent memory ranges from retrieval~\cite{lewis2020retrieval} to self-managed stores~\cite{packer2024memgpt,zhong2024memorybank}, and spoken history adds speech never meant for the assistant~\cite{contextdialog2025,xie2026voicemem}. Earlier agents named JARVIS orchestrate tools~\cite{shen2023hugginggpt} or play Minecraft~\cite{wang2023jarvis1}, but neither coordinates background work during a live conversation.

\section{Conclusion}
\label{sec:conclusion}

When an interactive model talks while a slow brain works in the background, each new turn can change what that work should do. \sysname gives this coordination a brain of its own, in which models make the judgments and code checks each one against the latest state of every task. With the front model and back end held fixed, this change alone brings large gains on every benchmark and front model we tested, and the design carries over to vision with little change. We expect the same split to help assistants in glasses, cars and robots keep talking naturally while they act.

{\raggedright\bibliography{ref}}

\clearpage
\appendix
\section{Implementation Details}
\label{app:impl}

This section collects the details of \sysname that the main text leaves out. Table~\ref{tab:instances} compares the two instances, and Table~\ref{tab:jev-decisions} lists every decision handed to \jev.

\begin{table}[!htbp]
\centering
\caption{The two instances of \sysname. Rows in the shared block use the same implementation in both.}
\label{tab:instances}
\tablefont
\setlength{\tabcolsep}{5pt}
\begin{tabularx}{\textwidth}{l >{\raggedright\arraybackslash}X >{\raggedright\arraybackslash}X}
\toprule
\tableheader & \textbf{\sysaudio} & \textbf{\sysomni} \\
\midrule
Front model & Realtime-Venus-Audio, 9B~\cite{venus2026} & Venus-Omni~\cite{venus2026} \\
\gate & Causal audio prefixes, three actions & Causal audio and screen features with dialogue context, adds a \act{SCENE} action for proactive turns \\
\brain & \jev, text & \jevomni, 12B, text and images \\
Perception state & ASR observer & ASR observer, \timeline and \belief \\
Memory & Profile and task records & Adds in-match events and game knowledge \\
\midrule
\tablegroup{3}{Shared implementation} \\
\multicolumn{3}{l}{\jev controller, \ledger, workers, \delivery and playback control} \\
\midrule
Main evidence & $\tau$-Voice, Full-Duplex-Bench v2 and v3, ContextDialog, race and control studies, daily use, PvZ, Luanti & Gate test set, visual and playback ablations, live CS2 and League of Legends sessions \\
\bottomrule
\end{tabularx}
\end{table}

\begin{table}[!htbp]
\centering
\caption{Decisions that \sysname hands to \jev. All of them use the typed interface, and none returns free text. The threshold is the confidence below which the runtime falls back to a safe default or asks for a second review.}
\label{tab:jev-decisions}
\tablefont
\setlength{\tabcolsep}{4pt}
\begin{tabularx}{\textwidth}{l >{\raggedright\arraybackslash}X l c}
\toprule
\tableheader \textbf{Decision} & \textbf{Options written by the runtime} & \textbf{Type} & \textbf{Threshold} \\
\midrule
\tablegroup{4}{Turns and tasks} \\
Turn route & answer fast, memory, worker, clarify, control, ignore, wait & Choice & 0.75 \\
Early preparation on partial speech & wait, prefetch memory, prefetch worker & Choice & 0.75 \\
Operation on an admitted turn & answer, memory, cancel, refine, status or resume on each live task version & Choice & 0.80 \\
Task control & status, refine, replace, cancel or resume on each task version, stop speaking & Choice & 0.80 \\
Batch interpretation & new work, refine, cancel, status, overview, memory, save, stop speech, ignore & Choice & 0.78 \\
Target of an operation & each live task by objective, or none & Choice & 0.80 \\
\midrule
\tablegroup{4}{Memory and delivery} \\
Durable memory write & yes or no, whether the speaker states a lasting fact about themselves & Choice & 0.80 \\
Memory selection & use or skip, for each retrieved record & Choice & -- \\
Relevance of a record & probability that a record bears on the request & Noul & -- \\
Delivery timing & deliver now or defer & Choice & -- \\
\midrule
\tablegroup{4}{Games and screens} \\
Game authorization & execute, advise, hand back to the human, unclear & Choice & 0.85 \\
Visual fact review (\sysomni) & probability that the frame supports a claim and that it is still fresh & Noul & 0.80 \\
Tactical advice (\sysomni) & tactical advice and social replies for the current game & Choice & 0.50 \\
\bottomrule
\end{tabularx}
\end{table}

\subsection{Jarvis-Audio}
\label{app:audio}

\paragraph{Gate and binding.}
The LoRA adapter of the \gate has rank 8 and scale 16 and consists of 216 tensors. The routing pass builds its own prefix, and every pass that produces text or speech runs with the adapter switched off. The endpointing of the ASR observer uses an ensemble of three Silero VAD~\cite{SileroVAD} instances and declares the end of an utterance after 1.1~s of silence. A \act{DELEGATE} event is bound to the most recent final transcript that started before the decision sample. Two adapter profiles share this design. A general profile serves daily use and all benchmarks, and a game profile serves sessions in which a game is explicitly connected.

\paragraph{Decision requests.}
When \jev classifies a turn as a batch, a planner splits it into contiguous spans copied verbatim from the transcript, and each span is interpreted on its own. Every planned operation passes the same check as a \jev choice. Its action must be one that the runtime offers, a refinement or cancellation must name an existing task, and the version must still be live when it is applied.

\paragraph{Ledger and workers.}
A task moves through eight states, and three of them, \state{CANCELLED}, \state{SUPERSEDED} and \state{FAILED}, are terminal. When the only change since a worker planned is newly heard but unconfirmed speech, a bounded check asks whether that speech is related to the task, and the plan proceeds only on a clear ``unrelated'' with no further change during the check. Clearing the history writes a time fence into the ledger, so an asynchronous write that started before the reset cannot bring an old task back afterwards. A worker receives the objective, the relevant history and memory, the public policy and the tool schemas of the domain. A result names its task and version, a spoken answer, the supporting evidence and any artifacts, and a request for a missing detail moves the task to \state{NEEDS\_INPUT}.

\paragraph{Delivery and memory.}
A routing decision that belonged to an older generation epoch is released back to the gate. A later request to hear a result again creates a fresh presentation in the current epoch. Each profile record has a kind, its text, the entities it mentions, the events it came from, a validity period, a status that is either active or superseded and an optional link to the record it replaces. An edited file is kept as a new revision in the task store. Clearing the history writes a fence, so an extraction that was still running cannot restore the cleared facts.

\subsection{Jarvis-Omni}
\label{app:omni}

\paragraph{Perception and decisions.}
The \timeline aligns its media tracks with a monotonic clock whose drift is estimated continuously. Some facts must come from the screen alone. A low-health reading, for example, is taken only from the HUD and never from the user saying ``I'm low''. \jevomni makes every judgment in \sysomni, from routing turns and binding them to tasks to reviewing visual claims and choosing tactical advice (Table~\ref{tab:jev-decisions}). It has a 256-way decision head and reads causal still frames and short audio windows. For the small and dense HUDs of games it adds light LoRA modules of about 8.5M parameters each, which read the HUD, keep champion identities consistent and detect whether the controlled character is alive. Open-ended visual descriptions are proposed by Gemma-4-12B-IT~\cite{gemma4} and must pass a separate review before the assistant relies on them.

\paragraph{Gate.}
The scene-aware gate is a rank-8 LoRA with 6,193,152 parameters, and the speech weights stay frozen. Its dialogue context includes earlier admitted user turns, the assistant audio that was actually played, the user profile and the task state. The adapter was trained with GRPO on the three native actions for two rounds of 3,000 steps, using 1,211 episodes built from 49 CS2 and League of Legends recordings and six synthetic speakers, split into 934 training, 181 development and 96 test episodes. Labels come from a teacher model with a blind second review. Table~\ref{tab:omni-gate} gives the result on the test set. In live sessions on the same 36 user recordings, all 18 replies are heard with the new adapter, against 13 with the previous one.

\begin{table}[!htbp]
\centering
\caption{The scene-aware gate of \sysomni on the fixed test set of 96 episodes, with 30 listening, 33 direct and 33 delegated targets. Early means acting before the reference time.}
\label{tab:omni-gate}
\tablefont
\setlength{\tabcolsep}{8pt}
\begin{tabular}{lccc}
\toprule
\tableheader \textbf{Control head} & \textbf{Correct} $\uparrow$ & \textbf{Missed} $\downarrow$ & \textbf{Early} \\
\midrule
Frozen Venus-Omni & 30 & 66 & 0 \\
Previous adapter & 67 & 22 & 1 \\
\oursrow \textbf{Contextual GRPO adapter} & \textbf{75} & \textbf{1} & 13 \\
\bottomrule
\end{tabular}
\end{table}

\paragraph{Memory on screen.}
\sysomni extends \memory with two layers. An in-match event memory records what happened during play, such as a kill, a suggestion or a correction, scoped to the user, the game, the match, the scene epoch and the round. Every event states whether it was observed, inferred or reported by the user, and whether it came from the screen or the microphone. Only an explicit correction from the user can supersede an event. A game-knowledge layer stores facts about a game version, each with a source URL and a revision. Events from earlier matches remain available but are labeled as belonging to another session, and a local preference such as ``stay quiet this round'' never becomes a lasting one. Recalled memory can be cited only as a past event, and recalled text can never re-enter the \timeline as if it had been seen.

\section{Training the Native Gate}
\label{app:gate}

All training of the \gate uses data we authored and synthesized ourselves.

\paragraph{Scenarios.}
We wrote 264 Chinese and English scenarios (142 and 122) that cover what a voice assistant hears in daily use. They include everyday questions and small talk, lookups that need tools, requests to read memory, changes and cancellations of running tasks, background speech, remarks addressed to other people, and discussion of games versus requests to act in them. Each scenario has a label from a written rubric, \act{DIRECT} for 76, \act{DELEGATE} for 156 and \act{LISTEN} for 32, and a context of no game, a PvZ session or a voxel sandbox (160, 52 and 52). Later rounds added three kinds of material. \emph{Continuity} scenarios cover 38 families of turns that only make sense with what came before, such as names, spelled identifiers, addresses, quantities, budgets and confirmations. \emph{Telephony} variants pass the speech through 8~kHz resampling, G.711 $\mu$-law coding and colored noise at 15~dB and 9~dB. \emph{Conversational} requests in 24 families, such as asking for a story, a joke, an explanation or some encouragement, teach the gate that a long answer is not the same as a task. Table~\ref{tab:gate-data} lists every data layer.

\paragraph{Speech, prefixes and splits.}
Every scenario is rendered with Qwen3-TTS~\cite{qwen3tts2026} in zero-shot mode from reference speakers and encoded by the streaming encoder of Venus-Audio. Prefixes that end in the silence before the utterance or in the middle of it are labeled \act{LISTEN}, and prefixes that end after it carry the rubric label. The final stage uses 654 episodes, 478 for training and 176 for development, which expand into 4,111 new causal prefixes and 1,056 replayed from earlier stages.

\begin{table}[!htbp]
\centering
\caption{Data layers of the \gate of \sysaudio. Later layers reuse earlier recordings under new channel conditions or contexts.}
\label{tab:gate-data}
\tablefont
\setlength{\tabcolsep}{5pt}
\begin{tabularx}{\textwidth}{l l X}
\toprule
\tableheader \textbf{Layer} & \textbf{Size and split} & \textbf{Content} \\
\midrule
Authored scenarios & 264, 144 train, 40 dev, 80 held out & Questions, small talk, tool needs, memory reads, task changes, background speech, game talk and game actions, in Chinese and English \\
Game round & 154, 82 train, 16 dev, 56 held out & Discussing a position, explaining it, authorizing an action and taking control back \\
Continuity & 38 families, 76 episodes & Names, spelled identifiers, addresses, quantities, budgets, edits, cancellations and confirmations, clean and over a telephone channel \\
Telephony & 246 episodes, 166 train, 80 dev & Quiet line and colored noise at 15~dB and 9~dB after 8~kHz resampling and $\mu$-law coding \\
Expanded telephony & 510 episodes, 382 train, 128 dev & 174 source scenarios under three channel conditions \\
Conversational requests & 24 families, 144 episodes & Stories, jokes, explanations, encouragement, side remarks and requests to wait, in three audio variants and two contexts \\
Final stage & 654 episodes, 478 train, 176 dev & Expanded telephony plus conversational requests, 618 used by the general profile \\
\bottomrule
\end{tabularx}
\end{table}

\paragraph{Objective.}
We optimize the policy of Equation~(\ref{eq:policy}) with GRPO~\cite{shao2024deepseekmath}. For each prefix $x$ we sample $G=8$ actions from a behavior distribution that mixes the current policy with a small uniform share,
\begin{equation}
q_\theta(a\mid x) = 0.8\,\pi_\theta(a\mid x) + 0.2/3 .
\end{equation}
A correct action earns a reward of $+2$ and a wrong one $-1$, and rewards are normalized within the group, $A_i = (r_i-\bar r)/(\mathrm{std}(r)+10^{-6})$. The loss combines a clipped policy-gradient term with a KL penalty toward the adapter at the start of the stage and a small entropy bonus,
\begin{equation}
\mathcal{L}(\theta) = -\,\mathbb{E}_i\Big[\min\big(\rho_i A_i,\ \mathrm{clip}(\rho_i,0.8,1.2)\,A_i\big)\Big] + \beta\, D_{\mathrm{KL}}\big(\pi_\theta\,\Vert\,\pi_{\mathrm{ref}}\big) - \lambda\, H(\pi_\theta),
\end{equation}
with $\rho_i = q_\theta(a_i\mid x)/q_{\mathrm{old}}(a_i\mid x)$, $\beta=0.03$ and $\lambda=0.002$. There is no value network and no token-level supervised loss. The KL anchor keeps the native distribution close to where each stage started, which protects behavior the data does not cover, and rewards on actions make trade-offs such as how early the gate may act easy to adjust. Table~\ref{tab:hparams} lists the settings for both instances.

\begin{table}[!htbp]
\centering
\caption{Training settings of the \gate for both instances.}
\label{tab:hparams}
\tablefont
\setlength{\tabcolsep}{8pt}
\adjustbox{max width=\textwidth}{%
\begin{tabular}{lcc}
\toprule
\tableheader \textbf{Setting} & \textbf{\sysaudio, final stage} & \textbf{\sysomni} \\
\midrule
Adapter & LoRA on q, v, o, rank 8, scale 16 & same \\
Trainable parameters & 6,193,152 & 6,193,152 \\
Frozen weights & all base weights & base, encoders and speech \\
Actions & \act{LISTEN}, \act{DIRECT}, \act{DELEGATE} & same three \\
Sampled actions per prefix & 8 & 8 \\
Reward & $+2$ correct, $-1$ wrong & correct versus wrong \\
Uniform exploration share & 0.2 & 0 \\
KL reference and coefficient & stage-initial adapter, 0.03 & frozen base, 0.003 \\
Entropy coefficient & 0.002 & -- \\
Sampling and evaluation temperature & 2.0 and 1.0 & 8.0 and 1.0 \\
Learning rate & $5\times10^{-6}$, AdamW, weight decay 0.01 & $1\times10^{-5}$ \\
Prefixes per step & 4, two new and two replayed & 3 \\
Training prefixes & 4,111 new and 1,056 replayed & 14,361 \\
Steps & 800 & two rounds of 3,000 \\
Selection & earlier dev must not drop & step 2,400 \\
\bottomrule
\end{tabular}%
}
\end{table}

\paragraph{Curriculum and selection.}
Training proceeds in four stages, each starting from the adapter of the previous one. The first stage learns causal timing on the authored scenarios. The second adds telephony, the third extends the scenario pool, and the fourth adds conversational requests. Every batch holds two prefixes from the new material and two replayed from the old, and half of the \act{LISTEN} slots go to complete speech that is not addressed to the assistant, which is the hardest kind of negative. The final stage runs 800 steps and takes 870~s. Checkpoints are evaluated every 100 steps, and a checkpoint is eligible only if its accuracy on the development set of the earlier stage does not drop. The game profile is chosen by interpolating between the adapters of the first two stages.

\begin{figure}[!htbp]
\centering
\includegraphics[width=\textwidth]{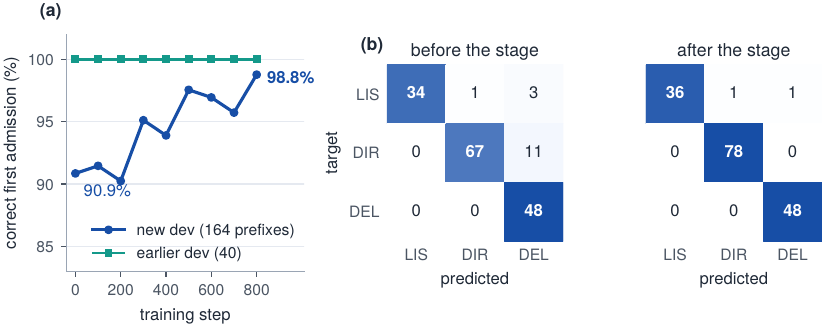}
\caption{Final training stage of the \gate. (a) Correct first admissions on the development sets, evaluated every 100 steps. (b) Confusion matrices on the new development set before and after the stage. Rows are targets and columns are predictions.}
\label{fig:gate-training}
\end{figure}

\paragraph{Results.}
Figure~\ref{fig:gate-training} shows the final stage. Accuracy on the new development set rises from 149 of 164 to 162 of 164 (90.9\% to 98.8\%), while the earlier development set stays at 40 of 40 throughout. Before this stage the gate sent 11 of 78 conversational requests to the slow brain, mostly requests for stories, jokes and encouragement that sound like work. After it, all 78 are answered directly. All 48 tasks are still delegated. Correct actions fire 0.16~s after the end of the request at the median.

\section{Evaluation Protocol}
\label{app:protocol}

\paragraph{Benchmarks.}
The 278 regular $\tau$-Voice tasks comprise 50 in airline, 114 in retail and 114 in telecom. FDB-v3 scores tool selection, argument accuracy, response quality and Pass@1, together with turn-taking behavior. For ContextDialog we fix 80 questions in advance, all four questions of 20 complete conversations from both official splits.

\paragraph{Front models and admission.}
MiniCPM-o~4.5 has no native function head, so a neutral controller admits requests after the final transcript. VoiceChat-11B has a native function head and a channel for tool results, and its own head decides when to call a tool. In every paired comparison the two harnesses share the same admission path. On Venus-Audio both use the trained \gate, and on the other two models both use the admission mechanism just described. Every harness uses a budget of 16,384 output tokens per task in addition to 12 tool rounds and 120~s.

\paragraph{Simulated users and judges.}
All benchmarks run locally with their official task sets, tools, environments and reward or scoring code. The simulated caller of $\tau$-Voice is driven by DeepSeek-V4-Flash and voiced by Qwen3-TTS under the official policies, noise conditions and 200~ms tick, and the examiner of FDB-v2 combines DeepSeek-V4-Flash, Parakeet-TDT~\cite{parakeet2025} and Qwen3-TTS. Judges use the official prompts and rubrics. Every system we run meets the same callers, examiners and judges, and audio reaches it as real-time PCM.

\paragraph{Baselines.}
QAA runs from its upstream code with its own gateway, task manager, backend runtime, announcements, cancellation and playback receipts. Our adapter supplies the duplex transport and a DeepSeek back end. Gander runs with its released weights, its lean orchestration mode and an official Whisper large-v3 recognizer, and a DeepSeek worker provider replaces its Codex back end with the same budget as the other harnesses. MoshiRAG runs on its native retrieval path. The official Venus harness runs with a DeepSeek executor.

\section{Additional Results}
\label{app:results}

\paragraph{Where the paired gains come from.}
Table~\ref{tab:fdb3-paired} breaks the paired FDB-v3 results down. With the same admissions on Venus-Audio, QAA produces an audible response in 68 of the 100 recordings and \sysname in 97. On VoiceChat argument accuracy rises from 0.470 to 0.583. On $\tau$-Voice, the hardest tasks require a write to the database, and QAA and Gander solve none of them. On FDB-v2 the largest gains appear in Entity tracking (1.14 points on Venus-Audio) and Safety (0.98 points). On Venus-Audio the mean reward rises by 0.62 on retail and 0.70 on telecom, with paired bootstrap intervals of $[0.54, 0.71]$ and $[0.61, 0.78]$.

\begin{table}[!htbp]
\centering
\caption{FDB-v3 quality and completion time with the same front model. Take-turn counts recordings with an audible response. Completion time is the official task-completion latency in seconds, mean with the median in parentheses.}
\label{tab:fdb3-paired}
\tablefont
\setlength{\tabcolsep}{6pt}
\begin{tabular}{llccccc}
\toprule
\tableheader \textbf{Front model} & \textbf{Harness} & \textbf{Take-turn}$\uparrow$ & \textbf{Tool F1}$\uparrow$ & \textbf{Arg.\ acc.}$\uparrow$ & \textbf{Resp.\ qual.}$\uparrow$ & \textbf{Completion (s)} \\
\midrule
\multirow{2}{*}{Venus-Audio} & QAA & 68 & 0.794 & 0.657 & 0.529 & 14.3 (14.2) \\
 & \cellcolor{oursRow}\textbf{\sysname} & \cellcolor{oursRow}\textbf{97} & \cellcolor{oursRow}\textbf{0.923} & \cellcolor{oursRow}\textbf{0.740} & \cellcolor{oursRow}\textbf{0.753} & \cellcolor{oursRow}15.0 (12.9) \\
\midrule
\multirow{2}{*}{MiniCPM-o 4.5} & QAA & 92 & 0.788 & 0.638 & 0.511 & 13.1 (13.1) \\
 & \cellcolor{oursRow}\textbf{\sysname} & \cellcolor{oursRow}\textbf{99} & \cellcolor{oursRow}\textbf{0.881} & \cellcolor{oursRow}\textbf{0.720} & \cellcolor{oursRow}\textbf{0.626} & \cellcolor{oursRow}14.0 (16.2) \\
\midrule
\multirow{2}{*}{VoiceChat-11B} & QAA & 96 & 0.772 & 0.470 & 0.396 & 22.6 (15.7) \\
 & \cellcolor{oursRow}\textbf{\sysname} & \cellcolor{oursRow}\textbf{97} & \cellcolor{oursRow}0.758 & \cellcolor{oursRow}\textbf{0.583} & \cellcolor{oursRow}\textbf{0.443} & \cellcolor{oursRow}17.2 (17.4) \\
\bottomrule
\end{tabular}
\end{table}

\paragraph{FDB-v3 by difficulty and disfluency.}
Table~\ref{tab:fdb3-breakdown} breaks down the published comparison. \sysaudio passes 56.7\% of the hard scenarios against 43.3\% for GPT-Realtime and 64.7\% of the recordings with self-corrections against 58.8\%.

\begin{table}[!htbp]
\centering
\caption{FDB-v3 Pass@1 (\%) by domain, difficulty and type of disfluency. Bold marks the best value in each column.}
\label{tab:fdb3-breakdown}
\tablefont
\setlength{\tabcolsep}{2.6pt}
\begin{tabular}{lcccc ccc ccccc}
\toprule
\tableheader & \multicolumn{4}{c}{\textbf{Domain}} & \multicolumn{3}{c}{\textbf{Difficulty}} & \multicolumn{5}{c}{\textbf{Disfluency}} \\
\tableheader \textbf{System} & Ecom. & Fin. & Hous. & Travel & Easy & Med. & Hard & Filler & Pause & Hesit. & False st. & Self-corr. \\
\cmidrule(lr){2-5} \cmidrule(lr){6-8} \cmidrule(lr){9-13}
GPT-Realtime & 55.2 & \textbf{96.0} & 30.8 & 60.0 & 75.0 & 58.8 & 43.3 & 62.1 & 55.6 & \textbf{70.0} & \textbf{66.7} & 58.8 \\
Gemini Live 2.5 & 48.3 & 76.0 & 23.1 & 50.0 & 66.7 & 50.0 & 26.7 & 62.1 & 44.4 & 60.0 & 41.7 & 47.1 \\
Gemini Live 3.1 & 44.8 & 92.0 & 26.9 & 55.0 & 69.4 & 58.8 & 30.0 & 58.6 & 50.0 & 60.0 & 58.3 & 35.3 \\
Grok & 41.4 & 76.0 & 11.5 & 45.0 & 58.3 & 47.1 & 20.0 & 48.3 & 33.3 & 50.0 & 58.3 & 29.4 \\
Ultravox & 34.5 & 68.0 & 19.2 & 45.0 & 55.6 & 38.2 & 26.7 & 41.4 & 33.3 & 50.0 & 25.0 & 35.3 \\
Cascaded & 41.4 & 80.0 & 19.2 & 40.0 & 63.9 & 44.1 & 23.3 & 44.8 & 44.4 & 60.0 & 50.0 & 17.6 \\
\midrule
\oursrow \textbf{\sysaudio} & \textbf{79.3} & 84.0 & \textbf{42.3} & \textbf{70.0} & \textbf{83.3} & \textbf{64.7} & \textbf{56.7} & \textbf{75.9} & \textbf{72.2} & \textbf{70.0} & 58.3 & \textbf{64.7} \\
\bottomrule
\end{tabular}
\end{table}

\begin{figure}[!htbp]
\centering
\includegraphics[width=\textwidth]{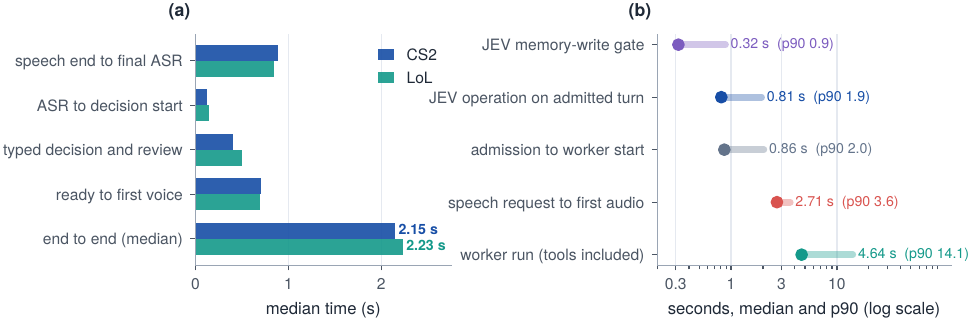}
\caption{Where the time goes. (a) The fast path of \sysomni in recorded CS2 and League of Legends sessions, with stage medians and the end-to-end median from the end of the user's speech to the first voiced word. (b) Stages of the slow brain of \sysaudio over 42 benchmark sessions.}
\label{fig:latency}
\end{figure}

\paragraph{Responsiveness.}
\label{sec:exp-latency}
Figure~\ref{fig:latency}(b) shows the stages of \sysaudio. \jev settles an admitted turn in 0.81~s at the median and the memory-write gate in 0.32~s, and a worker starts 0.86~s after admission. Most of the remaining time goes into the work itself, 4.6~s at the median for the tool loop and 2.7~s from a speech request to the first generated audio, while the fast brain keeps listening. With MiniCPM-o as the front end, \sysname starts speaking 3.1~s after the end of an FDB-v3 request at the median, slightly earlier than QAA on the same model at 3.5~s. Figure~\ref{fig:latency}(a) breaks down the fast path of \sysomni.

\end{document}